\input harvmac
\input amssym
\input epsf
\baselineskip 13pt

\def\p{\partial}
\def\Wc{\cal{W}}
\def\half{{1\over 2}}
\def\rar{\rightarrow}
\def\mw{{\cal{W}}}

\def\hsl{hs[$\lambda$]}
\def\sln{SL(N)}
\def\slt{SL(2)}
\def\slth{SL(3,R)}
\def\vphi{\varphi}
\def\vs{\vskip .1 in}
\def\a{\alpha}
\def\b{\beta}

\def\l{\lambda}

\def\ya{y_{\a}}

\def\za{z_{\a}}
\def\yt{\tilde{y}}

\def\yta{\tilde{y}_{\a}}
\def\zta{\tilde{z}_{\a}}
\def\ytb{\tilde{y}_{\beta}}

\def\yto{\tilde{y}_1}
\def\ytt{\tilde{y}_2}

\def\ep{e^{\rho}}

\def\pr{\p_{\rho}}

\def\pz{\p_z}
\def\pzb{\p_{\bar{z}}}

\def\emp{e^{-\rho}}

\def\mwil{${\cal{W}}_{\infty}[\l]$}

\def\zbar{\bar{z}}
\def\zb{\bar{z}} 

\def\psib{\overline{\psi}}
\def\mwil{\mw_{\infty}[\l]}
\def\pb{\overline{\p}}
\def\fsn{f^{s,n}_{\pm}(\l)}
\def\Ls{\Lambda^{(s)}}
\def\Ds{D^{(s)}}
\def\p{\partial}

\def\rt{\rightarrow}
\def\Oc{{\cal O}}

\def\zb{\overline{z}}

\def\psib{\overline{\psi}}

\def\Ct{\widetilde{C}}
\def\Ab{\overline{A}}

\def\zb{\overline{z}}

\def\Wc{{\cal W}}

\def\gb{\overline{g}}

\def\gt{\tilde{g}}

\def\phib{\overline{\phi}}

\def\yt{\tilde{y}}

\def\rar{\rightarrow}
\def\hsl{hs$[\lambda]$}
\def\Ocb{\overline{{\cal O}}}

\def\Ocb{\overline{\Oc}}

\def\phih{\widehat{\phi}}
\def\Ah{\widehat{A}}

\lref\CampoleoniZQ{
  A.~Campoleoni, S.~Fredenhagen, S.~Pfenninger and S.~Theisen,
  ``Asymptotic symmetries of three-dimensional gravity coupled to higher-spin
  fields,''
  JHEP {\bf 1011}, 007 (2010)
  [arXiv:1008.4744 [hep-th]].
}

\lref\BanadosGG{
  M.~Banados,
  ``Three-dimensional quantum geometry and black holes,''
  arXiv:hep-th/9901148.
}

\lref\BanadosNR{
  M.~Banados and R.~Caro,
  ``Holographic Ward identities: Examples from 2+1 gravity,''
  JHEP {\bf 0412}, 036 (2004)
  [arXiv:hep-th/0411060].
}

\lref\KrausVU{
  P.~Kraus, F.~Larsen and A.~Shah,
  ``Fundamental Strings, Holography, and Nonlinear Superconformal Algebras,''
  JHEP {\bf 0711}, 028 (2007)
  [arXiv:0708.1001 [hep-th]].
}

\lref\KrausNB{
  P.~Kraus and F.~Larsen,
  ``Partition functions and elliptic genera from supergravity,''
  JHEP {\bf 0701}, 002 (2007)
  [arXiv:hep-th/0607138].
}

\lref\DHokerHR{
  E.~D'Hoker, P.~Kraus and A.~Shah,
  arXiv:1012.5072 [hep-th].
}

\lref\FreedmanGP{
  D.~Z.~Freedman, S.~S.~Gubser, K.~Pilch and N.~P.~Warner,
  ``Renormalization group flows from holography supersymmetry and a c
  theorem,''
  Adv.\ Theor.\ Math.\ Phys.\  {\bf 3}, 363 (1999)
  [arXiv:hep-th/9904017].
}

\lref\BilalCF{
  A.~Bilal,
  ``W algebras from Chern-Simons theory,''
  Phys.\ Lett.\  B {\bf 267}, 487 (1991).
}

\lref\PopeCR{
  C.~N.~Pope and K.~S.~Stelle,
  ``SU(infinity), SU+(infinity) AND AREA PRESERVING ALGEBRAS,''
  Phys.\ Lett.\  B {\bf 226}, 257 (1989).
}

\lref\BergshoeffNS{
  E.~Bergshoeff, M.~P.~Blencowe and K.~S.~Stelle,
  ``AREA PRESERVING DIFFEOMORPHISMS AND HIGHER SPIN ALGEBRA,''
  Commun.\ Math.\ Phys.\  {\bf 128}, 213 (1990).
}

\lref\CampoleoniHG{
  A.~Campoleoni, S.~Fredenhagen and S.~Pfenninger,
  ``Asymptotic W-symmetries in three-dimensional higher-spin gauge theories,''
  arXiv:1107.0290 [hep-th].
}

\lref\HullKF{
  C.~M.~Hull,
  ``Lectures on W gravity, W geometry and W strings,''
  arXiv:hep-th/9302110.
}

\lref\EguchiSB{
  T.~Eguchi and H.~Ooguri,
  ``Conformal and Current Algebras on General Riemann Surface,''
  Nucl.\ Phys.\  B {\bf 282}, 308 (1987).
}

\lref\BergshoeffYD{
  E.~Bergshoeff, C.~N.~Pope, L.~J.~Romans, E.~Sezgin and X.~Shen,
  ``THE SUPER W(infinity) ALGEBRA,''
  Phys.\ Lett.\  B {\bf 245}, 447 (1990).
}

\lref\PopeKC{
  C.~N.~Pope, L.~J.~Romans and X.~Shen,
  ``A NEW HIGHER SPIN ALGEBRA AND THE LONE STAR PRODUCT,''
  Phys.\ Lett.\  B {\bf 242}, 401 (1990).}

\lref\WittenHC{
  E.~Witten,
  ``(2+1)-Dimensional Gravity as an Exactly Soluble System,''
  Nucl.\ Phys.\  B {\bf 311}, 46 (1988).
}

\lref\AchucarroVZ{
  A.~Achucarro and P.~K.~Townsend,
  ``A Chern-Simons Action for Three-Dimensional anti-De Sitter Supergravity
  Theories,''
  Phys.\ Lett.\  B {\bf 180}, 89 (1986).
}

\lref\BlencoweGJ{
  M.~P.~Blencowe,
  ``A Consistent Interacting Massless Higher Spin Field Theory In D = (2+1),''
Class.\ Quant.\ Grav.\  {\bf 6}, 443 (1989).
}

\lref\BrownNW{
  J.~D.~Brown and M.~Henneaux,
  ``Central Charges in the Canonical Realization of Asymptotic Symmetries: An
  Example from Three-Dimensional Gravity,''
  Commun.\ Math.\ Phys.\  {\bf 104}, 207 (1986).
}


\lref\HaggiManiRU{
  P.~Haggi-Mani and B.~Sundborg,
  ``Free large N supersymmetric Yang-Mills theory as a string theory,''
  JHEP {\bf 0004}, 031 (2000)
  [arXiv:hep-th/0002189].
}

\lref\GiombiYA{
  S.~Giombi, X.~Yin,
  ``On Higher Spin Gauge Theory and the Critical O(N) Model,''
[arXiv:1105.4011 [hep-th]].
}

\lref\KonsteinBI{
  S.~E.~Konstein, M.~A.~Vasiliev and V.~N.~Zaikin,
  ``Conformal higher spin currents in any dimension and AdS/CFT
  correspondence,''
  JHEP {\bf 0012}, 018 (2000)
  [arXiv:hep-th/0010239].
}

\lref\SundborgWP{
  B.~Sundborg,
  ``Stringy gravity, interacting tensionless strings and massless higher
  spins,''
  Nucl.\ Phys.\ Proc.\ Suppl.\  {\bf 102}, 113 (2001)
  [arXiv:hep-th/0103247].
}

\lref\MikhailovBP{
  A.~Mikhailov,
  ``Notes on higher spin symmetries,''
  arXiv:hep-th/0201019.
}

\lref\SezginRT{
  E.~Sezgin and P.~Sundell,
  ``Massless higher spins and holography,''
  Nucl.\ Phys.\  B {\bf 644}, 303 (2002)
  [Erratum-ibid.\  B {\bf 660}, 403 (2003)]
  [arXiv:hep-th/0205131].
}

\lref\KlebanovJA{
  I.~R.~Klebanov and A.~M.~Polyakov,
  ``AdS dual of the critical O(N) vector model,''
  Phys.\ Lett.\  B {\bf 550}, 213 (2002)
  [arXiv:hep-th/0210114].
}

\lref\GiombiWH{
  S.~Giombi and X.~Yin,
  ``Higher Spin Gauge Theory and Holography: The Three-Point Functions,''
  JHEP {\bf 1009}, 115 (2010)
  [arXiv:0912.3462 [hep-th]].
}

\lref\GiombiVG{
  S.~Giombi and X.~Yin,
  ``Higher Spins in AdS and Twistorial Holography,''$\quad \quad$
  arXiv:1004.3736 [hep-th].
}

\lref\HenneauxXG{
  M.~Henneaux and S.~J.~Rey,
  ``Nonlinear W(infinity) Algebra as Asymptotic Symmetry of Three-Dimensional
  Higher Spin Anti-de Sitter Gravity,''
  JHEP {\bf 1012}, 007 (2010)
  [arXiv:1008.4579 [hep-th]].
}

\lref\CampoleoniZQ{
  A.~Campoleoni, S.~Fredenhagen, S.~Pfenninger and S.~Theisen,
  ``Asymptotic symmetries of three-dimensional gravity coupled to higher-spin
  fields,''
  JHEP {\bf 1011}, 007 (2010)
  [arXiv:1008.4744 [hep-th]].
}

\lref\GaberdielAR{
  M.~R.~Gaberdiel, R.~Gopakumar and A.~Saha,
  ``Quantum W-symmetry in AdS$_3$,''
  JHEP {\bf 1102}, 004 (2011)
  [arXiv:1009.6087 [hep-th]].
}

\lref\GaberdielPZ{
  M.~R.~Gaberdiel and R.~Gopakumar,
  ``An AdS$_3$ Dual for Minimal Model CFTs,''
  arXiv:1011.2986 [hep-th].
}

\lref\DouglasRC{
  M.~R.~Douglas, L.~Mazzucato and S.~S.~Razamat,
  ``Holographic dual of free field theory,''
  arXiv:1011.4926 [hep-th].
}

\lref\CastroCE{
  A.~Castro, A.~Lepage-Jutier and A.~Maloney,
  ``Higher Spin Theories in AdS$_3$ and a Gravitational Exclusion Principle,''
  JHEP {\bf 1101}, 142 (2011)
  [arXiv:1012.0598 [hep-th]].
}

\lref\GaberdielWB{
  M.~R.~Gaberdiel and T.~Hartman,
  ``Symmetries of Holographic Minimal Models,''
  arXiv:1101.2910 [hep-th].
}

\lref\FradkinKS{
  E.~S.~Fradkin and M.~A.~Vasiliev,
  ``On the Gravitational Interaction of Massless Higher Spin Fields,''
  Phys.\ Lett.\  B {\bf 189}, 89 (1987).
}

\lref\FradkinQY{
  E.~S.~Fradkin and M.~A.~Vasiliev,
  ``Cubic Interaction in Extended Theories of Massless Higher Spin Fields,''
  Nucl.\ Phys.\  B {\bf 291}, 141 (1987).
}

\lref\VasilievTK{
  M.~A.~Vasiliev,
  ``Equations of motion for interacting massless fields of all spins in
  (3+1)-dimensions,''
{\it  In *Moscow 1990, Proceedings, Symmetries and algebraic structures in physics, pt. 1* 15-33}
}

\lref\VasilievAV{
  M.~A.~Vasiliev,
  ``More On Equations Of Motion For Interacting Massless Fields Of All Spins In
  (3+1)-Dimensions,''
  Phys.\ Lett.\  B {\bf 285}, 225 (1992).
}

\lref\BlencoweGJ{
  M.~P.~Blencowe,
  ``A Consistent Interacting Massless Higher Spin Field Theory In D = (2+1),''
  Class.\ Quant.\ Grav.\  {\bf 6}, 443 (1989).
}

\lref\BergshoeffNS{
  E.~Bergshoeff, M.~P.~Blencowe and K.~S.~Stelle,
  ``Area Preserving Diffeomorphisms And Higher Spin Algebra,''
  Commun.\ Math.\ Phys.\  {\bf 128}, 213 (1990).
}

\lref\ZamolodchikovWN{
  A.~B.~Zamolodchikov,
  ``Infinite Additional Symmetries In Two-Dimensional Conformal Quantum Field
  Theory,''
  Theor.\ Math.\ Phys.\  {\bf 65}, 1205 (1985)
  [Teor.\ Mat.\ Fiz.\  {\bf 65}, 347 (1985)].
}

\lref\KrausWN{
  P.~Kraus,
  ``Lectures on black holes and the AdS(3)/CFT(2) correspondence,''
  Lect.\ Notes Phys.\  {\bf 755}, 193 (2008)
  [arXiv:hep-th/0609074].
}

\lref\FronsdalRB{
  C.~Fronsdal,
  ``Massless Fields With Integer Spin,''
  Phys.\ Rev.\  D {\bf 18}, 3624 (1978).
}

\lref\DidenkoTD{
  V.~E.~Didenko and M.~A.~Vasiliev,
  ``Static BPS black hole in 4d higher-spin gauge theory,''
  Phys.\ Lett.\  B {\bf 682}, 305 (2009)
  [arXiv:0906.3898 []].
}
\lref\BakasRY{
  I.~Bakas, E.~Kiritsis,
  ``BOSONIC REALIZATION OF A UNIVERSAL W ALGEBRA AND Z(infinity) PARAFERMIONS,''
Nucl.\ Phys.\  {\bf B343}, 185-204 (1990).
}

\lref\KochCY{
  R.~d.~M.~Koch, A.~Jevicki, K.~Jin and J.~P.~Rodrigues,
  ``$AdS_4/CFT_3$ Construction from Collective Fields,''
  Phys.\ Rev.\  D {\bf 83}, 025006 (2011)
  [arXiv:1008.0633].
}


\lref\GutperleKF{
  M.~Gutperle and P.~Kraus,
  ``Higher Spin Black Holes,''
  JHEP {\bf 1105}, 022 (2011)
  [arXiv:1103.4304 [hep-th]].
}

\lref\AhnPV{
  C.~Ahn,
  ``The Large N 't Hooft Limit of Coset Minimal Models,''
  arXiv:1106.0351 [hep-th].
}

\lref\GaberdielZW{
  M.~R.~Gaberdiel, R.~Gopakumar, T.~Hartman and S.~Raju,
  ``Partition Functions of Holographic Minimal Models,''
  arXiv:1106.1897 [hep-th].
}

\lref\BershadskyBG{
  M.~Bershadsky,
  ``Conformal field theories via Hamiltonian reduction,''
  Commun.\ Math.\ Phys.\  {\bf 139}, 71 (1991).
}

\lref\PolyakovDM{
  A.~M.~Polyakov,
  ``Gauge Transformations and Diffeomorphisms,''
  Int.\ J.\ Mod.\ Phys.\  A {\bf 5}, 833 (1990).
}

\lref\BilalCF{
  A.~Bilal,
  ``W algebras from Chern-Simons theory,''
  Phys.\ Lett.\  B {\bf 267}, 487 (1991).
}

\lref\DynkinUM{
  E.~B.~Dynkin,
  ``Semisimple subalgebras of semisimple Lie algebras,''
  Trans.\ Am.\ Math.\ Soc.\  {\bf 6}, 111 (1957).
}

\lref\BaisBS{
  F.~A.~Bais, T.~Tjin and P.~van Driel,
  ``Covariantly coupled chiral algebras,''
  Nucl.\ Phys.\  B {\bf 357}, 632 (1991).
}

\lref\ForgacsAC{
  P.~Forgacs, A.~Wipf, J.~Balog, L.~Feher and L.~O'Raifeartaigh,
  ``Liouville and Toda Theories as Conformally Reduced WZNW Theories,''
  Phys.\ Lett.\  B {\bf 227}, 214 (1989).
}

\lref\deBoerIZ{
  J.~de Boer and T.~Tjin,
  ``The Relation between quantum W algebras and Lie algebras,''
  Commun.\ Math.\ Phys.\  {\bf 160}, 317 (1994)
  [arXiv:hep-th/9302006].
}

\lref\ChangMZ{
  C.~M.~Chang and X.~Yin,
  ``Higher Spin Gravity with Matter in $AdS_3$ and Its CFT Dual,''
  arXiv:1106.2580 [hep-th].
}

\lref\MaldacenaRE{
  J.~M.~Maldacena,
  ``The Large N limit of superconformal field theories and supergravity,''
  Adv.\ Theor.\ Math.\ Phys.\  {\bf 2}, 231 (1998)
  [Int.\ J.\ Theor.\ Phys.\  {\bf 38}, 1113 (1999)]
  [arXiv:hep-th/9711200].
}
\lref\FigueroaOCV{
  J.~M.~Figueroa-O'Farrill, J.~Mas, E.~Ramos,
  ``A One parameter family of Hamiltonian structures for the KP hierarchy and a continuous deformation of the nonlinear W(KP) algebra,''
Commun.\ Math.\ Phys.\  {\bf 158}, 17-44 (1993).
[hep-th/9207092].
}
\lref\WittenQJ{
  E.~Witten,
  ``Anti-de Sitter space and holography,''
  Adv.\ Theor.\ Math.\ Phys.\  {\bf 2}, 253 (1998)
  [arXiv:hep-th/9802150].
}

\lref\GubserBC{
  S.~S.~Gubser, I.~R.~Klebanov and A.~M.~Polyakov,
  ``Gauge theory correlators from noncritical string theory,''
  Phys.\ Lett.\  B {\bf 428}, 105 (1998)
  [arXiv:hep-th/9802109].
}

\lref\JevickiSS{
  A.~Jevicki, K.~Jin and Q.~Ye,
  ``Collective Dipole Model of AdS/CFT and Higher Spin Gravity,''
  arXiv:1106.3983 [hep-th].
}

\lref\WaldNT{
  R.~M.~Wald,
  ``Black hole entropy is the Noether charge,''
  Phys.\ Rev.\  D {\bf 48}, 3427 (1993)
  [arXiv:gr-qc/9307038].
}

\lref\ProkushkinBQ{
  S.~F.~Prokushkin and M.~A.~Vasiliev,
  ``Higher spin gauge interactions for massive matter fields in 3-D AdS
  space-time,''
  Nucl.\ Phys.\  B {\bf 545}, 385 (1999)
  [arXiv:hep-th/9806236].
}

\lref\ProkushkinVN{
  S.~Prokushkin and M.~A.~Vasiliev,
  ``3-d higher spin gauge theories with matter,''
  arXiv:hep-th/9812242.
}

\lref\tmg{
  B.~Chen, J.~Long and J.~b.~Wu,
  ``Spin-3 Topological Massive Gravity,''
  arXiv:1106.5141 [hep-th].}

\lref\tmgg{
  A.~Bagchi, S.~Lal, A.~Saha and B.~Sahoo,
  ``Topologically Massive Higher Spin Gravity,''
  arXiv:1107.0915 [hep-th].}

\lref\gab{
  M.~R.~Gaberdiel and T.~Hartman,
  ``Symmetries of Holographic Minimal Models,''
  JHEP {\bf 1105}, 031 (2011)
  [arXiv:1101.2910 [hep-th]].}

\lref\cfp{
  A.~Campoleoni, S.~Fredenhagen and S.~Pfenninger,
 ``Asymptotic W-symmetries in three-dimensional higher-spin gauge theories,''
  arXiv:1107.0290 [hep-th].}

\lref\prs{
  C.~N.~Pope, L.~J.~Romans and X.~Shen,
  ``W(infinity) and the Racah-Wigner Algebra,''
  Nucl.\ Phys.\  B {\bf 339}, 191 (1990).}

\lref\cacc{
  S.~L.~Cacciatori, M.~M.~Caldarelli, A.~Giacomini, D.~Klemm and D.~S.~Mansi,
  ``Chern-Simons formulation of three-dimensional gravity with torsion and
  nonmetricity,''
  J.\ Geom.\ Phys.\  {\bf 56}, 2523 (2006)
  [arXiv:hep-th/0507200].}

\lref\VasilievBA{
  M.~A.~Vasiliev,
  ``Higher spin gauge theories: Star product and AdS space,''
  arXiv:hep-th/9910096.
}

\lref\BekaertVH{
  X.~Bekaert, S.~Cnockaert, C.~Iazeolla and M.~A.~Vasiliev,
  ``Nonlinear higher spin theories in various dimensions,''
  arXiv:hep-th/0503128.
}

\lref\SundborgWP{
  B.~Sundborg,
  ``Stringy gravity, interacting tensionless strings and massless higher
  spins,''
  Nucl.\ Phys.\ Proc.\ Suppl.\  {\bf 102}, 113 (2001)
  [arXiv:hep-th/0103247].
}

\lref\DasVW{
  S.~R.~Das and A.~Jevicki,
  ``Large N collective fields and holography,''
  Phys.\ Rev.\  D {\bf 68}, 044011 (2003)
  [arXiv:hep-th/0304093].
}

\lref\VasilievEN{
  M.~A.~Vasiliev,
  ``Consistent equation for interacting gauge fields of all spins in
  (3+1)-dimensions,''
  Phys.\ Lett.\  B {\bf 243}, 378 (1990).
}

\lref\VasilievCM{
  M.~A.~Vasiliev,
  ``Closed equations for interacting gauge fields of all spins,''
  JETP Lett.\  {\bf 51}, 503 (1990)
  [Pisma Zh.\ Eksp.\ Teor.\ Fiz.\  {\bf 51}, 446 (1990)].
}

\lref\ChangMZ{
  C.~M.~Chang and X.~Yin,
  ``Higher Spin Gravity with Matter in AdS$_3$ and Its CFT Dual,''
  arXiv:1106.2580 [hep-th].
}

\lref\AmmonNK{
  M.~Ammon, M.~Gutperle, P.~Kraus and E.~Perlmutter,
  ``Spacetime Geometry in Higher Spin Gravity,''
  arXiv:1106.4788 [hep-th].
}

\lref\StromingerEQ{
  A.~Strominger,
  ``Black hole entropy from near horizon microstates,''
  JHEP {\bf 9802}, 009 (1998)
  [arXiv:hep-th/9712251].
}

\lref\BanadosWN{
  M.~Banados, C.~Teitelboim and J.~Zanelli,
  ``The Black hole in three-dimensional space-time,''
  Phys.\ Rev.\ Lett.\  {\bf 69}, 1849 (1992)
  [arXiv:hep-th/9204099].
}

\lref\CampoleoniHG{
  A.~Campoleoni, S.~Fredenhagen and S.~Pfenninger,
  ``Asymptotic W-symmetries in three-dimensional higher-spin gauge theories,''
  arXiv:1107.0290 [hep-th].
}

\lref\WaldNT{
  R.~M.~Wald,
  ``Black hole entropy is the Noether charge,''
  Phys.\ Rev.\  D {\bf 48}, 3427 (1993)
  [arXiv:gr-qc/9307038].
}

\lref\GibbonsUE{
  G.~W.~Gibbons and S.~W.~Hawking,
  ``Action Integrals and Partition Functions in Quantum Gravity,''
  Phys.\ Rev.\  D {\bf 15}, 2752 (1977).
}

\lref\PopeEW{
  C.~N.~Pope, L.~J.~Romans and X.~Shen,
  ``The Complete Structure of W(Infinity),''
  Phys.\ Lett.\  B {\bf 236}, 173 (1990).
}

\lref\BakasRY{
  I.~Bakas and E.~Kiritsis,
  ``Bosonic realization of a universal W algebra and Z(infinity) Parafermions,''
  Nucl.\ Phys.\  B {\bf 343}, 185 (1990)
  [Erratum-ibid.\  B {\bf 350}, 512 (1991)].
}

\lref\HullSA{
  C.~M.~Hull,
  ``W gravity anomalies 1: Induced quantum W gravity,''
  Nucl.\ Phys.\  B {\bf 367}, 731 (1991).
}

\lref\HullKF{
  C.~M.~Hull,
  ``Lectures on W gravity, W geometry and W strings,''
  arXiv:hep-th/9302110.
}

\lref\EguchiSB{
  T.~Eguchi and H.~Ooguri,
  ``Conformal and Current Algebras on General Riemann Surface,''
  Nucl.\ Phys.\  B {\bf 282}, 308 (1987).
}

\lref\Iazeolla{
  C.~Iazeolla, P.~Sundell,
  ``Families of exact solutions to Vasiliev's 4D equations with spherical, cylindrical and biaxial symmetry,''
  [arXiv:1107.1217 [hep-th]].}

\lref\FreedmanTZ{
  D.~Z.~Freedman, S.~D.~Mathur, A.~Matusis, L.~Rastelli,
  ``Correlation functions in the CFT(d) / AdS(d+1) correspondence,''
Nucl.\ Phys.\  {\bf B546}, 96-118 (1999).
[hep-th/9804058].
}

\lref\HansenWU{
  J.~Hansen, P.~Kraus,
  ``Generating charge from diffeomorphisms,''
JHEP {\bf 0612}, 009 (2006).
[hep-th/0606230].
}

\lref\KrausWN{
  P.~Kraus,
  ``Lectures on black holes and the AdS(3) / CFT(2) correspondence,''
Lect.\ Notes Phys.\  {\bf 755}, 193-247 (2008).
[hep-th/0609074].
}

\lref\KrausDS{
  P.~Kraus, E.~Perlmutter,
  ``Partition functions of higher spin black holes and their CFT duals,''
[arXiv:1108.2567 [hep-th]].
}

\lref\GaberdielNT{
  M.~R.~Gaberdiel and C.~Vollenweider,
  ``Minimal Model Holography for SO(2N),''
  arXiv:1106.2634 [hep-th].
}

\lref\AhnBY{
  C.~Ahn,
  ``The Coset Spin-4 Casimir Operator and Its Three-Point Functions with Scalars,''
[arXiv:1111.0091 [hep-th]].
}

\lref\CastroFM{
  A.~Castro, E.~Hijano, A.~Lepage-Jutier and A.~Maloney,
  ``Black Holes and Singularity Resolution in Higher Spin Gravity,''
  arXiv:1110.4117 [hep-th].
}

\lref\TanTJ{
  H.~S.~Tan,
  ``Aspects of Three-dimensional Spin-4 Gravity,''
  arXiv:1111.2834 [hep-th].
}

\lref\KrausNB{
  P.~Kraus, F.~Larsen,
  ``Partition functions and elliptic genera from supergravity,''
JHEP {\bf 0701}, 002 (2007).
[hep-th/0607138].
}

\lref\PetkouZZ{
  A.~C.~Petkou,
  ``Evaluating the AdS dual of the critical O(N) vector model,''
JHEP {\bf 0303}, 049 (2003).
[hep-th/0302063].
}

\lref\LeighGK{
  R.~G.~Leigh, A.~C.~Petkou,
  ``Holography of the N=1 higher spin theory on AdS(4),''
JHEP {\bf 0306}, 011 (2003).
[hep-th/0304217].
}

\lref\SezginPT{
  E.~Sezgin, P.~Sundell,
  ``Holography in 4D (super) higher spin theories and a test via cubic scalar couplings,''
JHEP {\bf 0507}, 044 (2005).
[hep-th/0305040].
}

\Title{\vbox{\baselineskip14pt
}} {\vbox{\centerline {Scalar fields and three-point functions}
\medskip\vbox{\centerline {in $D=3$ higher spin gravity}}}}
\centerline{Martin Ammon, Per
Kraus and Eric Perlmutter\foot{ammon@physics.ucla.edu, pkraus@ucla.edu, perl@physics.ucla.edu}}
\bigskip
\centerline{\it{Department of Physics and Astronomy}}
\centerline{${}$\it{University of California, Los Angeles, CA 90095, USA}}

\baselineskip14pt

\vskip .3in

\centerline{\bf Abstract}
\vskip.2cm
We compute boundary  three-point functions involving two scalars and a gauge field of arbitrary spin in the AdS vacuum of Vasiliev's higher spin gravity, for any deformation parameter $\lambda$. In the process, we develop tools for extracting scalar field equations in arbitrary higher spin backgrounds.  We work in the context of \hsl$\oplus$\hsl\ Chern-Simons theory coupled to scalar fields, and make efficient use of  the associative lone-star product underlying the \hsl\ algebra. Our results for the  correlators
precisely match expectations from CFT; in particular they match those of any CFT with $\mw_{\infty}[\l]$ symmetry at large central charge,  and with primary operators dual to the scalar fields. As this is expected to include the `t Hooft limit of the $W_N$ minimal model CFT, our results serve as further evidence of the conjectured AdS/CFT duality between these two theories.

\Date{November  2011}
\baselineskip13pt

\listtoc \writetoc

\newsec{Introduction}

Higher spin gravity in AdS has been receiving increased attention in recent years for its promise in shedding light on certain big questions in string theory.
For one, strings propagating in an AdS spacetime of string scale curvature are expected to be described by some theory of interacting higher spins \refs{\HaggiManiRU,\SundborgWP,\SezginRT}; while the precise theory remains unknown, the theories developed by Vasiliev and collaborators \refs{\FronsdalRB,\FradkinKS,\FradkinQY,\VasilievEN, \VasilievCM,\VasilievAV} appear to be, at the very  least, sophisticated toy models of such a theory, with a large gauge symmetry and nonlocal dynamics. For another, the higher spin AdS/CFT dualities known thus far involve  vector-like boundary theories, that is, CFTs whose central charges scale as $N$ rather than $N^2$. The tractability of these vector models encourages the optimistic view that such duality conjectures may be derivable, allowing us to peer inside the black box of AdS/CFT.

The prototype duality proposal is that of Klebanov and Polyakov \KlebanovJA , who conjectured a duality between (a subsector of) Vasiliev's theory of higher spins on AdS$_4$ and the large $N$ limit of the 2+1-dimensional O(N) vector model. This proposal was studied early on in \refs{\PetkouZZ,\LeighGK,\SezginPT }, as well as in the  recent papers \refs{\GiombiWH,\GiombiVG,\GiombiYA,\KochCY,\JevickiSS}.  Through a combination of direct computation of bulk three-point functions, generalization to $n$-point functions, and arguments regarding higher spin symmetry breaking and AdS boundary conditions, these works provide a powerful body of evidence that the duality is correct, at least as far as the bulk theory is understood.

An AdS$_3$ analog of the Klebanov-Polyakov conjecture is the proposed duality \GaberdielPZ\ between Vasiliev gravity in D=3 \ProkushkinBQ\ coupled to a pair of complex scalar fields and a 't Hooft limit of the $W_N$ minimal model CFT, which has a coset representation
\eqn\inb{ {SU(N)_k \oplus SU(N)_1 \over SU(N)_{k+1}} }
The 't Hooft limit is defined as
\eqn\inc{ N, k \rt \infty~,\quad    \lambda \equiv {N \over k+N}~~~{\rm fixed} }

There is already  quite a bit of evidence for this conjecture.  This includes matches between bulk and boundary global symmetries \GaberdielWB; the bulk 1-loop determinant and the large $N$ CFT partition function \GaberdielZW ; a higher spin black hole partition function at high temperature and the same quantity in the CFT \KrausDS ; and scalar-scalar-higher spin three-point functions at $\l=1/2$ and $s=2,3,4$, in the 't Hooft limit \refs{\ChangMZ,\AhnBY}.

It is this last piece of evidence upon which we build in the present work, with the aim of drawing closer to a perturbative proof of the conjecture a la \GiombiVG. There are two immediate ways we would like to generalize previous calculations of three-point correlators. The bulk computation of \ChangMZ\ was in the ``undeformed'' Vasiliev theory, that is, at $\l=1/2$ (or $\nu=0$ in the language of \ProkushkinBQ), and we want to extend this to  arbitrary $\l$. The minimal model computations of \refs{\ChangMZ,\AhnBY} only treat the low spins, and we want to compute for arbitrary spin.

In this paper, we succeed in computing the bulk correlators for arbitrary spin and arbitrary $\lambda$, and show that they match precisely with CFT expectations.  On the CFT side we proceed under the assumption that  the theory has $\Wc_\infty[\lambda]$ symmetry.  A separate question, not addressed here, is whether this is indeed true of the minimal model CFTs in the 't Hooft limit; evidence in favor appears in \refs{\GaberdielWB,\GaberdielZW,\AhnBY}. Along the way, we present techniques for writing generalized bulk scalar wave equations in arbitrary on-shell higher spin backgrounds, which have interesting applications beyond the present context.

We now state our result for the three-point correlators. First, in the deformed bulk theory, we calculate the three-point function of two scalar fields and a higher spin field of arbitrary spin $s$. Recall that there are two complex scalar fields in the bulk, each with $m^2=\l^2-1$. Taking one of them to be dual to an operator $\Oc$ and its complex conjugate $\overline{\Oc}$, our result for the three-point function in terms of the scalar-scalar two-point function is
\eqn\ind{\eqalign{\langle{\Oc}_{\pm}(z_1)\overline{\Oc}_{\pm}(z_2)J^{(s)}(z_3)\rangle&= {(-1)^{s-1}\over 2\pi }{\Gamma(s)^2\over\Gamma(2s-1)}{\Gamma(s\pm\l)\over\Gamma(1\pm\l)}\left({z_{12}\over z_{13}z_{23}}\right)^s\langle{\Oc}_{\pm}(z_1)\overline{\Oc}_{\pm}(z_2)\rangle \cr}}
where the subscript denotes standard (+) or alternative ($-$) quantization of the scalar. The same correlator for the other scalar field, dual to an operator $\widetilde{\Oc}$ and its complex conjugate $\overline{\widetilde{\Oc}}$, is identical to \ind\ but absent the $(-1)^s$ prefactor. In the context of the duality \GaberdielPZ, these operators should be assigned to take opposite quantization to one another.

Let us now mention a few of the main insights that allowed us to compute these correlators relatively easily.   First, it is well known that  if the scalar fields are  set to zero the Vasiliev theory is equivalent to a Chern-Simons theory with gauge algebra \hsl$\oplus$\hsl.   To compute correlators of the type \ind\ we need to couple a free scalar field to this theory.  The general rules for incorporating scalar fields into the higher spin theory are complicated but known (see \ProkushkinBQ\ and appendix A).  However, free scalar field equations can be derived from the elegant equation\foot{A second type of scalar, dual to the $\widetilde{\Oc}$ operators noted above, obeys the same equation but with $A$ and $\Ab$ interchanged.}
\eqn\inda{dC + A\star C - C\star \Ab =0}
where $(A,\Ab)$ denote the \hsl$\oplus$\hsl\ gauge fields. As we will explain, $C$ is a ``master field" that takes values in the Lie algebra \hsl\ supplemented with an identity element, and the scalar is the part of $C$ proportional to the identity.    This equation reduces to the Klein-Gordon equation for a scalar of mass $m^2 = \lambda^2-1$ when evaluated in AdS, and in general gives the coupling of higher spin gauge fields to the linearized scalar. An important role is played by the star-product appearing in \inda, which is an associative multiplication known as the ``lone-star product" \prs.  In the original paper \ProkushkinBQ\ this product is realized in terms of the Moyal product applied to  symmetric polynomials of  ``deformed oscillators."  However, the deformed oscillator approach is rather inconvenient for present purposes, as one has to deal with the tedious procedure of resymmetrizing strings of oscillators.  By contrast, the lone-star product gives us a closed form expression for the multiplication rules, and turns out to be much simpler to work with.

The second key insight is to put the higher spin gauge invariance at center stage. To compute three-point correlators of the type \ind\ we need to solve \inda\ in the presence of flat connections $A$ and $\Ab$ representing higher spin gauge fields with prescribed asymptotics.   Such flat connections can be generated by gauge transformations.  Therefore, if we start from a solution for the free scalar in AdS$_3$ and then act with the gauge transformation, we generate a new scalar solution in the presence of the higher spin gauge fields.  In this way, rather than having to first work out the perturbed scalar equation and then solve it, we can generate the solution in one step, which is a huge simplification.

On the CFT side, our starting point is the assumption that in the 't Hooft limit the $W_N$ coset CFT has $\mwil$ global symmetry. While this is unproven, it is a prerequisite for the duality to hold in the pure gauge sector. Previous calculations \refs{\ChangMZ,\AhnBY} took the tack of computing the $s=2,3,4$ correlators \ind\ at finite $N$ in the CFT, and taking the 't Hooft limit afterwards; this serves as good evidence that $\mwil$ really does emerge in the 't Hooft limit, but the complications of finite $N$ are not required if one wants to ask questions about the scalar sector. Thus, for the purposes of calculating the correlator \ind, we believe that the most direct strategy begins with the assumed symmetry of the 't Hooft limit, and asks to what extent the representation theory of $\mwil$ --- in particular, of its large $k$ wedge subalgebra \hsl\ --- fixes the correlator.  Generalizing results in \GaberdielWB, we show that, in fact, it is enough to reproduce the result \ind\ and the accompanying result for the second scalar, providing perfect agreement with the bulk.

Our results for  bulk and boundary correlators reduce to  previous computations \refs{\ChangMZ,\AhnBY} in the appropriate limits. In all, we find this to be a significant step toward verifying the duality proposal \GaberdielPZ.

Our techniques will also have application to computing other correlation functions in these theories.   For instance, there is by now a good understanding of higher spin black holes in D=3 \refs{\GutperleKF,\AmmonNK,\KrausDS,\CastroFM,\TanTJ}; in particular their entropy is known to match that of the dual CFT in the high temperature limit \KrausDS.  Using the results of this paper, it is now quite feasible to compute scalar correlators in the background of a higher spin black hole and compare with CFT.

The remainder of this paper is organized as follows.  In section 2 we go through the main steps involved in deriving the equation \inda\ from the general formulation of the Vasiliev theory. Further details are provided in appendix A.  In section 3 we show how to work out the explicit form of the scalar wave equation in the presence of higher spin gauge fields.  In section 4, which is the core of the paper, we show how to use gauge invariance to generate solutions of the scalar wave equations, and then read off the desired correlation functions.  In section 5 we compute these correlators on the CFT side under the assumption of $\Wc_\infty[\lambda]$ symmetry, and demonstrate perfect agreement with the bulk.  We conclude with some comments in section 6.  Appendix B presents some evidence for the isomorphism between the lone-star product and the Moyal product acting on deformed oscillators, in Appendix C we derive a result needed in the text, and Appendix D provides some useful explicit expressions for comparison with a formula derived in section 4.

\newsec{Matter fields in Vasiliev gravity}

We begin with a review of the formulation of $D=3$ higher spin gravity due to Vasiliev and collaborators, as presented in \ProkushkinBQ.
 We first recall how to write the gauge sector of this theory as a \hsl$\oplus$\hsl\ Chern-Simons theory, and then show how to introduce linearized scalar fields  in the Chern-Simons language. Seeing as we will not need all of the details of the theory's construction, we present an abridged discussion; the reader who would prefer not to take anything on faith is referred to appendix A.

Vasiliev gravity contains one higher spin gauge field for each integer spin $s\geq 2$, coupled to some number of matter multiplets. There are various ingredients, foremost among them a set of ``master fields:'' a spacetime 1-form $W = W_\nu dx^\nu$ as well as spacetime 0-forms $B$ and $S_\alpha.$ Besides the spacetime coordinates $x,$ the generating functions $W, B$ and $S_\alpha$ also depend on auxiliary bosonic twistor variables\foot{The $\za,\ya$ are sometimes referred to as "oscillators."} $\za,\ya$ where $\alpha=1,2,$ as well as on two pairs of Clifford elements: $\psi_{1,2}$, and $k,\rho$. That is,
\eqn\inta{ \{ \psi_i, \psi_j \} = 2 \delta_{ij}~, ~~ k\rho = - \rho k~, ~~  k^2 = \rho^2 =1}
Whereas $\psi_{1,2}$ commute with all other auxiliary variables, $k$ and $\rho$ have the properties
\eqn\intb{\eqalign{
k y_\alpha &= - y_\alpha k, \quad k z_\alpha = - z_\alpha k \cr
\rho y_\alpha &=y_\alpha \rho, \quad\ \ \rho z_\alpha = z_\alpha \rho}}

Roughly speaking, $W$ encodes the gauge sector, $B$ parameterizes the AdS vacua of the theory and is used to introduce propagating matter fields, and $S_{\a}$ ensures that the theory has the correct internal symmetries. The elements $\{\za,\ya,k,\rho\}$ are ingredients in the realization of these symmetries, and the $\psi_i$ are required only when writing down solutions.

Twistor indices are raised  and lowered by the rank two antisymmetric tensor $\epsilon_{\alpha\beta}$:
\eqn\eqIIg{ z^\alpha = \epsilon^{\alpha\beta} z_\beta \, , \qquad\quad z_{\alpha} = z^{\beta} \epsilon_{\beta\alpha}}
where we use the convention, following  \ProkushkinBQ, $\epsilon^{12}=\epsilon_{12}=1.$
Functions of the twistors $\za,\ya$ are multiplied by the Moyal product:
\eqn\intc{ f(z,y) \star g(z,y) = {1\over{(2\pi)^2}} \int d^2u \int d^2v \ e^{i u_\alpha v^\alpha} f(z+u,y+u) g(z-v, y+v)}

We can combine the oscillators and various  other ingredients to construct two sets of so-called ``deformed'' oscillators, denoted $\zta,\yta$. The $\yta$, which will be more important for our work here, obey deformed oscillator star-commutation relations
\eqn\intd{\eqalign{ [\tilde{y}_\alpha, \tilde{y}_\beta]_\star &= 2i \epsilon_{\alpha\beta} (1+\nu k) \cr}}
These deformed oscillators give rise to the higher spin algebra \hsl\ as follows.
Define elements of the algebra to consist of symmetrized, positive even-degree polynomials in $\tilde{y}_\alpha$.  Multiplying these elements using \intd, and projecting onto $k=\pm 1$, the commutation relations are those of \hsl, with $\lambda = {1\over 2}(1\mp \nu)$.\foot{More precisely, we consider polynomials of degree two and higher, since the constant term star-commutes with everything. We also note that one can enlarge \hsl\ by including odd powers of the $\yta$ in the polynomials, though our interests here are in purely bosonic Vasiliev theory.}

This deformed oscillator algebra plays a central role in the study of AdS vacua in Vasiliev theory.  It emerges dynamically from the field equations of the full nonlinear system of higher spins. The parameter $\nu$ encodes the deformation, and in the event that $\nu=0$, the theory is said to be undeformed. $\nu$ also plays two other important roles: it parameterizes a family of inequivalent AdS vacua, and sets the mass of any scalar fields that we introduce to the theory. These connections stem from the structure of the field equations, themselves tightly constrained by higher spin gauge invariance.

To see how $\nu$ appears in connection with the AdS vacua, we examine the field equations. There are five equations in terms of the master fields $W,B$ and $S_{\a}$, and we present two of them here (the others are written in the appendix):
\eqn\intf{\eqalign{
dW &= W  \wedge\star W \cr
dB &= W \star B - B \star W \cr
}}
The equations that we have omitted all depend on $S_\alpha$.  At the order to which we will be working in this paper --- namely, linear in the scalar fields --- the entire effect of $S_\alpha$ is that it forces $W$ and $B$ to be independent of $k$ and  $\tilde{z}_\alpha$.   With this in hand, we can proceed by focussing on \intf.

We first consider solutions with vanishing scalar field.  This corresponds to taking a constant background value for $B$,
\eqn\inth{B=\nu}
where the constant $\nu$ is fixed by the omitted $S_\alpha$ equations to be the same parameter as appears in the deformed oscillator expressions.

Now, the first equation in \intf\ can be written as a flatness condition for two Chern-Simons gauge fields, each taking values in the Lie algebra \hsl. In order to see this we introduce gauge fields $A$ and $\Ab$ by
\eqn\inti{ W = - {\cal P}_+  A - {\cal P}_- \Ab}
where $A$ and $\Ab$ are functions of $\tilde{y}_\alpha$ and the spacetime coordinates $x^\mu$. Here we have introduced the projection operators
\eqn\intj{ {\cal P}_\pm = {{1\pm\psi_1}\over{2}}}
obeying
\eqn\intk{ {\cal P}_\pm \psi_1 = \psi_1 {\cal P}_\pm = \pm {\cal P}_\pm, \qquad {\cal P}_\pm \psi_2 = \psi_2 {\cal P}_\mp}
Plugging \inti\ into \intf\ yields
\eqn\intl{\eqalign{  dA + A\wedge \star A &=0 \cr  d\Ab + \Ab \wedge \star \Ab &=0 }}
From our earlier remarks, we see that if $A$ and $\Ab$ are taken to be polynomials of positive even degree in symmetrized products of $\yt_\alpha$, then \intl\ are equivalent to the field equations of \hsl$\oplus$\hsl\ Chern-Simons theory. Since SL(2) is a subalgebra of \hsl, this theory includes ordinary Einstein gravity with a negative comological constant as a consistent truncation.

Before introducing the scalar fields, let us say a bit more about \hsl.
The \hsl\ Lie algebra is spanned by generators labeled by a spin index $s$  and a mode index $m$. We use the notation of \GaberdielWB, in which a generator is represented as
\eqn\ba{V^s_m~, ~~ s \geq 2~, ~~ |m|<s}
The commutation relations are
\eqn\bb{[V^s_m,V^t_n] = \sum_{u=2,4,6,...}^{s+t-|s-t|-1}g^{st}_u(m,n;\l)V^{s+t-u}_{m+n}}
with structure constants defined in appendix B. The generators with $s=2$ form an \slt\ subalgebra, and the remaining generators transform simply under the adjoint \slt\ action as
\eqn\be{[V^2_m,V^t_n] = (m(t-1)-n)V^t_{m+n}}
The parameter $\l$ can be mapped to the parameters of the oscillator formulation as
\eqn\bba{\l={1-\nu k\over 2}}
where $k=\pm 1$.
When $\l=1/2$, the theory is undeformed and this algebra is isomorphic to hs(1,1) \BlencoweGJ.

To summarize what we have found so far, the gauge sector of the Vasiliev theory boils down to \hsl$\oplus$\hsl\ Chern-Simons theory.
This sector of the theory has no propagating degrees of freedom.

To introduce propagating scalar fields we study a linearized fluctuation of $B$ around its vacuum value \inth,
\eqn\intm{B = \nu + {\cal C}(x,\psi_i,\yt_\alpha) }
The bosonic field ${\cal C}$ is taken  to have an expansion in even-degree symmetrized products of the deformed oscillators $\yt_\alpha$. The lowest term in the expansion, with no deformed oscillators, will be identified with the physical scalar field. ${\cal C}$ obeys
\eqn\intn{\eqalign{ d {\cal C} - W \star {\cal C} + {\cal C} \star W &=0 }}
We decompose ${\cal C}$ as
\eqn\intna{ {\cal C} = {\cal P}_+ \psi_2 C(x,\yt_\alpha) + {\cal P}_- \psi_2 \tilde{C}(x,\yt_\alpha)  }
Plugging into \intn\ we find
\eqn\intnb{\eqalign{  dC+ A\star C - C\star \Ab & =0 \cr
 d\Ct + \Ab \star \Ct- \Ct\star A   & =0 }}
As shown in the next section, expanded around AdS each of these equations reduces to the Klein-Gordon equation for a scalar field of mass $m^2 = \lambda^2-1$.  More generally, these equations capture the interaction of the linearized scalars with an arbitrary higher spin background. For example, they can be used to study the propagation of a scalar field in the higher spin black hole of \KrausDS.

We note that the two equations \intnb\ are related by $A \leftrightarrow \Ab$, $C\leftrightarrow \Ct$.  This is interpreted as  a ``charge conjugation" operation that flips the sign of all odd spin tensor gauge fields.  Another notable feature is that the equations \intnb\ are only sensible for $A$ and $\Ab$ on-shell, i.e. satisfying equation \intl.  This can be seen by taking $d$ of these equations; if the connections are not flat this leads to extra constraints on $C$ and $\Ct$ with no interpretation in terms of propagating scalar fields.

Recapping, we have now reduced the system of equations down to \intl\ and \intnb.  These equations describe the propagation of linearized scalar fields in an arbitrary on-shell higher spin background.  They do not capture backreaction of the scalar on the higher spin fields, or self-interactions among the scalars.   Neither of these effects is needed for the computation of the three-point correlators herein.

We now turn to solving these equations, and introducing efficient tools for this purpose. We focus on the  $C$ equation; results for $\Ct$ then follow by charge conjugation.

\newsec{Generalized Klein-Gordon equations in higher spin backgrounds}

Starting from
\eqn\gaa{dC +A\star C - C\star \Ab=0}
for some higher spin background determined by \hsl-valued connections $(A,\Ab)$, we want to extract the generalized Klein-Gordon equation hiding within.

As we said, in the traditional formalism of bosonic Vasiliev theory, the master field $C$ is expanded in deformed oscillators $\yta$ as
\eqn\ga{C = C^1_0 + C^{\a\b}\yta\ytb + C^{\a\b\sigma\l}\yta\ytb\yt_{\sigma}\yt_{\l}+\ldots}
where the star product is implied and all components of $C$ are symmetric in twistor indices. This separates the components of the master field into the physical scalar field, which is the lowest component $C^1_0$, and the remaining components related on-shell to $C^1_0$ by derivatives. Plugging \ga\ into \gaa\ leads, after much work, to the scalar equations.

The most tedious part of this computation is multiplying the deformed oscillators.  We need to take a pair of symmetrized combinations of oscillators, multiply them, and then resymmetrize using \intd.  Rather than carrying out this procedure each time, it would be much more convenient if we had a closed-form expression for the multiplication rules.  Recall that the Lie algebra obtained via star-commutation of these elements is \hsl. Now, underlying \hsl\ is an associative product, under which the \hsl\ Lie bracket becomes the commutator \bb.  This ``lone-star product" is defined as

\eqn\gb{V^s_m \star V^t_n \equiv \half \sum_{u=1,2,3,...}^{s+t-|s-t|-1}g^{st}_u(m,n;\l)V^{s+t-u}_{m+n}}
One recovers \bb\ upon using the fact that
\eqn\gbab{g^{st}_u(m,n;\l) = (-1)^{u+1}g^{ts}_u(n,m;\l)}
This was originally presented in an early paper on $W_{\infty}$ algebras \prs, and used more recently by two of the present authors \KrausDS\ to compute black hole partition functions in \hsl\ gravity.

It is then very natural to suspect an isomorphism between the product rules for symmetrized oscillator combinations and those of the lone-star product. In appendix B we present strong evidence at low spins that, indeed, the lone-star product acting on \hsl\ generators is isomorphic to the product involving the deformed oscillators, with a specific identification between generators and oscillator polynomials.

Proceeding under this assumption, which will be well justified by the consistency of all results, provides a major technical simplification.
One trades the tedious symmetrization procedure for the known and easily manipulated \hsl\ structure constants.

In this language, we expand the master field $C$ as follows:
\eqn\gd{C = \sum_{s=1}^{\infty}\sum_{|m|<s}C^s_mV^s_m}
This maps to \ga\ under the identifications
\eqn\ge{C^s_m \sim C^{\a_1\a_2\cdots\a_{2s-2}}}
with the index $m$ related to the number of oscillators $\yto$ versus $\ytt$ as $2m=N_1-N_2$. (For more details see, appendix B.) 
The functions $C^s_m$ are functions of spacetime coordinates $x$, and the auxiliary tensor structure has been absorbed into $(s,m)$.  The gauge fields are expanded similarly,
\eqn\gda{\eqalign{A &= \sum_{s=2}^{\infty}\sum_{|m|<s}A^s_mV^s_m~,\quad  \Ab = \sum_{s=2}^{\infty}\sum_{|m|<s}\Ab^s_mV^s_m} }
The lowest scalar  component, $C^1_0$, will be our physical scalar field, and the remaining $C^s_m$ will be related to  derivatives of $C^1_0$. Upon plugging into \gaa\ we can obtain the equation of motion for $C^1_0$.\vs

Let us begin by solving the equation \gaa\ in AdS; aside from being an instructive exercise, this will lay the foundation for what follows. The same computation has been performed in various places (e.g. \refs{\ProkushkinBQ,\ChangMZ}) in the oscillator language. Because there are no higher spin fields turned on in the vacuum, $C^1_0$ should obey the ordinary Klein-Gordon equation and the remaining components should be fixed in terms of $C^1_0$. In addition, previous literature on the subject tells us that the scalar mass squared is $m^2=\l^2-1$.   An attractive feature of the theory is that the value of the mass is fixed by the gauge algebra.

\subsec{AdS: Recovering the Klein-Gordon equation}
The AdS connection is constructed out of the spin-2 generators alone, namely those forming an \slt\ subalgebra of \hsl. We work in Euclidean signature in Fefferman-Graham gauge, with radial coordinate $\rho$ and boundary coordinates $(z,\zbar)$. The connection is
\eqn\gg{\eqalign{&A  = e^{\rho}V^2_1dz + V^2_0d\rho\cr
&\Ab = e^{\rho}V^2_{-1}d\zbar - V^2_0d\rho\cr}}
giving rise to a metric\foot{In this work, we will not need the prescription to pass from Chern-Simons to metric language; suffice it to say that in writing this metric, we have chosen a particular normalization of the \hsl\ trace. See \KrausDS\ for conventions used here.}
\eqn\ghh{ds^2 = d\rho^2 + e^{2\rho}dzd\zbar}
and vanishing higher spin fields.

Due to the simplicity of this connection --- the small number of generators, their low spin and the symmetric appearance of $A$ and $\Ab$ ---  the $C$ equations \gaa\ are simple to write. Decomposing along both spacetime and internal \hsl\ space, and using the lone-star product, one finds\foot{$V^s_{m,x^{\mu}}$ is a short-hand notation for the component along $V^s_m dx^{\mu}$.}
\eqn\gi{\eqalign{V^s_{m,\rho}:&\quad \pr C^s_m +2C^{s-1}_m+C^{s+1}_mg^{(s+1)2}_3(m,0)=0\cr
V^s_{m,z}:&\quad \p C^s_m + \ep\left(C^{s-1}_{m-1}+\half g^{2s}_2(1,m-1)C^s_{m-1} + \half g^{2(s+1)}_3(1,m-1)C^{s+1}_{m-1}\right)=0\cr
V^s_{m,\zbar}:&\quad \pb C^s_m - \ep\left(C^{s-1}_{m+1}-\half g^{2s}_2(-1,m+1)C^s_{m+1} + \half g^{2(s+1)}_3(-1,m+1)C^{s+1}_{m+1}\right)=0\cr}}
where $|m|< s$ and $\p=\pz, \pb=\pzb$. (Here and henceforth, we suppress the $\l$-dependence of the structure constants $g^{st}_u(m,n)$.)

It is now easy to obtain the Klein-Gordon equation. Writing out a handful of equations at $s=1,2$, one finds four that form a closed set for components $\{C^1_0, C^2_0, C^3_0, C^2_1\}$:
\eqn\gk{\eqalign{V^1_{0,\rho}: \quad &\p_{\rho}C^1_0+C^2_0\cdot{\l^2-1\over 6}=0\cr
V^1_{0,\zbar}: \quad &\pb C^1_0+\ep C^2_1\cdot{\l^2-1\over 6}=0\cr
V^2_{1,z}: \quad &\p C^2_1+\ep C^1_0 +{\ep\over 2} C^2_0-\ep C^3_0\cdot {\l^2-4\over 30}=0\cr
V^2_{0,\rho}: \quad &\p_{\rho}C^2_0+2C^1_0+C^3_0\cdot {2(\l^2-4)\over 15}=0\cr}}
Solving for the higher components and plugging back in yields
\eqn\gl{\left[\p^2_{\rho} + 2\p_{\rho}+4e^{-2\rho}\p\pb -(\l^2-1)\right]C^1_0=0}
This is the Klein-Gordon equation in the background \ghh\ with the correct scalar mass.

In order for the entire set of unfolded equations to be consistent, all components of $C$ must have a smooth solution in terms of $C^1_0$.\foot{Upon solving for the $C^s_m$ in terms of $C^1_0$, one observes the following pole structure in the $\l$ plane:
\eqn\sumf{C^s_m \sim (\cdots)\prod_{p=1}^{s-1}(\l^2-p^2)^{-1}}
This is evident in \gk, for example. These are not problematic for $p\neq 1$, as \hsl\ degenerates to \sln\ at integer values of $\l\geq 2$, and the spin $s>\l$ fields do not exist. The singularity at $\l=1$ has a different role, but is a reflection of the fact that hs[1], in the absence of a rescaling of generators, becomes similarly degenerate as many structure constants vanish.} We delay presentation of the solution for the full master field $C$ in AdS until section 4, where we will need it to compute the three-point functions.  There, we will also show that for any connection related to AdS by a non-singular gauge transformation, the linearized matter equations \gaa\ admit a consistent solution for the full master field $C$. \vs

Before moving on to higher spin deformations of AdS, let us elucidate the structure of \gi\ and present a systematic strategy for isolating the minimal set of equations needed to solve for $C^1_0$. We wish to highlight a special type of component of $C$, namely those which are of the form $C^{m+1}_{\pm m}$ and hence have the smallest possible spin for fixed mode $m$: $C^1_0, C^2_{\pm1}$, etc. We call these ``minimal'' components.

Starting with the $V^s_{m,\rho}$ equations, it is clear that for fixed mode $m$, one can solve these recursively for all non-minimal components in terms of $C^{m+1}_{\pm m}$ and $\rho$ derivatives thereof. This is a consequence of being in Fefferman-Graham gauge, whereby $A_{\rho}=-\Ab_{\rho}=V^2_0$, and we will remain in this gauge throughout this paper. Having solved for the minimal components $C^{m+1}_{\pm m}$, one should view the $V^s_{m,z}$ and $V^s_{m,\zbar}$ equations as determining these in terms of $C^1_0$ and $(z,\zbar)$ derivatives thereof.

This reveals a useful strategy for extracting the smallest possible set of equations to obtain the scalar equation in any background. We think of the $\rho$ equations as implicitly solved. Then, along $(z,\zbar)$, one need only keep track of the mode indices appearing in any given equation, and one need only look at equations along minimal directions.

Let us demonstrate with the AdS connection \gg.  We use the following heuristic for which components appear in which equations (structure constants implied): for modes $m=0,1,2$,

\eqn\gm{\eqalign{&\vdots\quad\quad\quad\quad\quad\quad\quad\quad\quad\quad\quad\vdots\cr
V_{0,\zbar}\sim~ & \pb C_0 +  C_{1}~,\quad V_{0,z}\sim~ \p C_0 +  C_{-1}\cr
V_{1,\zbar}\sim~& \pb C_1 +  C_{2}~,\quad
V_{1,z}\sim~ \p C_1 +  C_{0}\cr
V_{2,\zbar}\sim~& \pb C_2 +  C_{3}~,\quad
V_{2,z}\sim~ \p C_2 +  C_{1}\cr
&\vdots\quad\quad\quad\quad\quad\quad\quad\quad\quad\quad\quad\vdots}}
The equations $V^1_{0,\zbar},V^2_{1,z}$ form a closed set among components with $m=0,1$, and so will be enough, along with whatever $\rho$ equations we need, to find the Klein-Gordon equation. This is exactly what we presented in \gk.
\subsec{Chiral higher spin deformations of AdS}

To warm up to the higher spin connections we will ultimately consider, we present the simplest possible higher spin deformation of AdS: a constant, chiral spin-3 deformation,
\eqn\cha{\eqalign{A &= e^{\rho}V^2_{1}dz-\eta e^{2\rho}V^3_{2}d\zbar + V^2_0 d\rho\cr
\Ab &= e^{\rho} V^2_{-1}d\zbar - V^2_0 d\rho\cr}}
From the point of view of the boundary CFT, this corresponds to adding a dimension-3 operator to the CFT, with constant coupling $\eta$.

Using our mnemonic of the previous subsection, we make quick work of this connection. Again writing the $C$ master field equation in spacetime and gauge components, we show some of the $\zbar$ equations:
\eqn\chb{\eqalign{V_{-1,\zbar}\sim~ & \pb C_{-1} +  C_{0}+\eta C_{-3}\cr
V_{0,\zbar}\sim~ & \pb C_0 +  C_{1}+\eta C_{-2}\cr
V_{1,\zbar}\sim~& \pb C_1 +  C_{2}+\eta C_{-1}\cr
V_{2,\zbar}\sim~& \pb C_2 +  C_{3}+\eta C_0\cr}}
The components along $z$ are unchanged from the AdS case. Reinserting the spin indices,  the minimal closed set of equations is $\{V^1_{0,z}, V^2_{1,z}, V^3_{2,z}, V^2_{1,\zbar}\}$, along with any $\rho$ equations necessary. Solving this system gives the following equation for $C^1_0$:
\eqn\chc{\left[\p_{\rho}^2 + 2\p_{\rho}+4e^{-2\rho}(\p\pb-\eta \p^3) -(\l^2-1)\right]C^1_0=0}
There is now a three-derivative term, as one would expect from dimensional analysis.

This can be extended to a chiral spin-$s$ deformation: for a connection
 \eqn\chd{\eqalign{A &= e^{\rho}V^2_{1}dz-\eta e^{(s-1)\rho}V^s_{s-1}d\zbar + V^2_0 d\rho\cr
\Ab &= e^{\rho} V^2_{-1}d\zbar - V^2_0 d\rho\cr}}
the generalized Klein-Gordon equation is
\eqn\che{\left[\p_{\rho}^2 + 2\p_{\rho}+4e^{-2\rho}(\p\pb+\eta (-\p)^s) -(\l^2-1)\right]C^1_0=0}

This example nicely captures the primary general feature of higher spin deformations:  higher derivative terms enter the generalized wave equation.

One can extend these methods to any connection --- black holes or RG flows, for instance --- although the difficulty in solving the resulting set of equations increases rather quickly with the number of generators. We now study a slightly more complicated connection, relevant for computation of correlation functions in section 4.

\subsec{Higher spin currents in AdS}
Starting from \cha, we wish to allow $\eta$ to have arbitrary dependence on $(z,\zbar)$.\foot{When writing the functional dependence of fields and operators on $(z,\zbar)$, we temporarily use the notation $z\equiv(z,\zbar)$.} This will act as a source for spin-$3$ charge $J^{(3)}$, enabling us to compute the correlator $\langle {\Oc}(z_1)\overline{\Oc}(z_2)J^{(3)}(z_3)\rangle$, where $\Oc$ is a scalar primary dual to $C^1_0$. To do so, we will need to obtain the scalar equation to linear order in the spin-3 source (which we now label $\mu^{(3)}$).

Previous work on higher spin gravity \refs{\CampoleoniZQ,\GutperleKF}, following earlier work in pure gravity \BanadosNR, laid out a dictionary for relating sources and charges to components of the Chern-Simons connection, and we will apply and recapitulate those techniques here.

The following is a flat connection, to linear order in the source $\mu^{(3)}(z)$:
\eqn\chg{\eqalign{A_z &= e^{\rho}V^2_1 + {1\over B^{(3)}}J^{(3)}(z) e^{-2\rho}V^3_{-2}\cr
A_{\zbar} &=-\sum_{n=0}^{4}{1\over n!}((-\p)^n\mu^{(3)}(z))e^{(2-n)\rho}V^3_{2-n}\cr
\overline{A}_{\zbar}&=e^{\rho}V^2_{-1}\cr}}
along with the usual $A_{\rho} = -\Ab_{\rho}=V^2_0$,
subject to
\eqn\chh{\pb J^{(3)}(z)=-{B^{(3)}\over 4!}\p^5\mu^{(3)}(z)}

This is the same connection as first presented in \GutperleKF, now embedded in \hsl\ instead of \slth\ and with vanishing stress tensor. The leading term in $A_{\zbar}$, namely $-\mu^{(3)} e^{2\rho}V^3_2$, is the source, dual to the charge term ${1\over B^{(3)}}J^{(3)} e^{-2\rho}V^3_{-2}$ in $A_z$. The remaining terms in $A_{\zbar}$ are required for flatness. We have included a normalization constant $B^{(3)}$ in the definition of the current. While its actual value is unimportant for the calculations in this paper\foot{For an explicit formula for this coefficient for any spin, see \GaberdielWB, equation (A.4), where $B^{(s)} = -{k\over 2\pi}N_s$ in their notation.}, we include it to stress that we are inserting the factor
\eqn\chhb{e^{\int\! d^2z ~\mu^{(3)} J^{(3)} } }
in the CFT path integral. Writing the connection with unit coefficient (up to a sign) for the chemical potential $\mu^{(3)}$ then fixes the other free coefficient, and $B^{(3)}$ is fixed by the Ward identity \chh, equivalently by the OPE
\eqn\chha{J^{(3)}(z)J^{(3)}(0)\sim {5B^{(3)}\over 2\pi}{1\over z^6}}

Passing to the metric-like formulation of the spin-3 field using $\vphi_{\mu\nu\sigma} \sim\Tr ( e_{(\mu} e_\nu e_{\gamma)})$, one can explicitly check that this connection turns on various components of $\vphi_{\mu\nu\sigma}$: for instance, the component
\eqn\chia{\vphi_{\zbar\zbar\zbar} \sim \mu^{(3)} e^{4\rho}\Tr(V^3_2V^2_{-1}V^2_{-1})}
makes it clear that we have turned on a spin-3 source that grows toward the boundary.

Following our prior method, one finds that the following set of equations forms the minimal closed set needed to determine $C^1_0$:
\eqn\chii{V^2_{1,z}~, ~~ V^1_{0,z}~, ~~ V^2_{-1,z}~, ~~ V^3_{-2,z}~, ~~ V^2_{1,\zbar}~, ~~ V^1_{0,\zbar}~, ~~ V^2_{-1,\zbar}}
As always, these should be accompanied by some number of $V^s_{m,\rho}$ equations required to eliminate non-minimal components of $C$. Solving these perturbatively, one finds the following scalar equation to linear order in $\mu^{(3)}\equiv \mu$:
\eqn\chj{\left(\square_{KG} + \square_{\mu}\right)C^1_0=0}
where
\eqn\chll{\eqalign{\square_{KG} &= \pr^2+2\pr+4e^{-2\rho}\p\pb-(\l^2-1)\cr
\square_{\mu}&= -{1\over 6}e^{-6\rho}\left(\p^5\mu\pb^2+\p^4\mu\p\pb^2\right)+{4\over B^{(3)}}J^{(3)}e^{-4\rho}\pb^3\cr
&+{1\over 3}e^{-4\rho}\left(\p^4\mu\pb -\p^4\mu\pb\pr  + \p^3\mu\p\pb -\p^3\mu\p\pb\pr \right)\cr
&-{1\over 6}e^{-2\rho}\Big(3\p^3\mu\pr^2 -(\l^2-1)\p^3\mu +3\p^2\mu\p\pr^2-12\p^2\mu \p\pr  -(\l^2-13)\p^2\mu\p \cr
&+36\p\mu\p^2-12\p\mu\p^2\pr +24\mu\p^3 \Big)\cr}}
This equation may be used to infer the cubic vertex between two scalars and the metric-like spin-3 field. On the other hand, it does not clearly suggest what the analogous equation would look like for higher spin sources. What is clear is that the number of terms will grow with the spin.

To solve this using standard AdS/CFT methodology, we write
\eqn\chm{C^1_0(\rho,z;z_1) = G_{b\p}(\rho,z;z_1) + \phi_{\mu}(\rho,z;z_1)}
where the subscript denotes the order in $\mu$. The first term is the bulk-to-boundary propagator of a scalar in AdS with $m^2=\l^2-1$, and with standard (+) or alternative $(-$) quantization:
\eqn\chma{G_{b\p}(\rho,z;z_1)  = \pm{\l\over\pi}\left({e^{-\rho}\over e^{-2\rho}+|z-z_1|^2}\right)^{1\pm\l}}
The solution to \chj\ to linear order in $\mu$ is
\eqn\chn{\phi_{\mu}(\rho,z;z_1) = -\int d^2z'd\rho'e^{2\rho'}G_{bb}(\rho,z;\rho',z')\square_{\mu}'G_{b\p}(\rho',z';z_1)}
where $G_{bb}(\rho,z;\rho',z')$ is the bulk-to-bulk propagator obeying
\eqn\chpp{\square_{KG}G_{bb}(\rho,z;\rho',z') = e^{-2\rho}\delta(\rho-\rho')\delta^{(2)}(z-z') }
By judicious use of  integration by parts, one can reduce the integral to a boundary term and read off the three-point function. This was the strategy employed in \ChangMZ. Instead, we will step back and discuss a simpler method that makes full use of higher spin gauge invariance from the outset.  With this approach we bypass the need to first find the modified scalar equation and then solve it, and instead contruct the needed solution directly.

\newsec{Three-point correlators from the bulk}

We now turn to the main focus of this paper: the efficient computation of three-point correlation functions involving two scalar operators and one higher spin current.  Our basic observation is that starting from the solution for a free scalar field in AdS we can generate a new solution by performing a higher spin gauge transformation.  This essentially reduces the whole problem to determining how the scalar field transforms under the gauge transformation.

\subsec{Spin-1 example}

To illustrate our general approach in the simplest context, in this section we compute the three-point function of two scalar operators and a spin-1 current. Rather than working in the Vasiliev theory, here we take
the bulk action  to be a complex scalar field of mass $m^2 = \lambda^2-1$ coupled to a U(1) Chern-Simons gauge field,
\eqn\pkaa{ S =  {k\over 4\pi} \int\! A\wedge dA + {1\over 2} \int\! d^3x \sqrt{g} \Big( |D^\mu \phi|^2 + (\lambda^2-1)|\phi|^2 \Big)}
with $D_\mu = \p_\mu +A_\mu$.  To compute the correlator $\langle \Oc(z_1) \Ocb(z_2) J^{(1)}(z_3)\rangle$ we proceed as follows.  We insert delta function sources at $z_2$ and $z_3$ by imposing the following  asymptotic behavior on the scalar and gauge field\foot{The reason for the hats will be clear momentarily. Also, note here that we are using standard quantization for the scalar.}
\eqn\pkab{ \phih(\rho,z) \sim  \mu_\phi \delta^{(2)}(z-z_2)e^{-(1-\lambda) \rho}~,\quad \Ah_{\zb}(\rho,z)  \sim \mu_A \delta^{(2)}(z-z_3)~,\quad \rho \rt \infty}
This form for the gauge field corresponds to a source for the boundary current, as explained in \refs{\HansenWU,\KrausNB,\KrausWN}.
We then need to find the order $\mu_\phi \mu_A$ contribution to the vev
$\Oc(z_1)$, which  also can  be read off from the scalar field asymptotics,
\eqn\pkac{ \phih(\rho,z) \sim  {\Oc(z) \over B_\phi}e^{-(1+\lambda)\phi}~,\quad \rho\rt \infty,\quad z \neq z_{2,3}}
We will keep the constant $B_\phi$ unspecified, though we note that a consistent holographic dictionary fixes $B_\phi =2\lambda$ (for $\lambda \neq 0$)  \FreedmanTZ.  The three-point function is then given by
\eqn\pkad{ \Oc(z_1) = \mu_\phi \mu_A  \langle \Oc(z_1) \Ocb(z_z) J^{(1)}(z_3)\rangle + \cdots }
where $\cdots$ denote terms of other order in $\mu_{\phi,A}$.

Since the gauge field has no propagating degrees of freedom we can generate the required solution by a gauge transformation.   In particular, we start by solving for the scalar field with $A=0$, using the bulk-boundary propagator
\eqn\pkae{ \phi(\rho,z) =  \int \! d^2z' G_{b\p}(\rho,z;z_2)\phi_-(z')  }
This solution corresponds to an arbitrary source $ \phi_-(z)$.
To generate the desired gauge field solution we apply a gauge transformation
\eqn\pkaf{  A_\mu = \p_\mu \Lambda~,\quad \Lambda(z) = {\mu_A\over 2\pi} {1\over z-z_3} }
where we use the formula $\p_{\zb} \left({1 \over z}\right) =2\pi \delta^{(2)}(z)$.    The gauge transformation acts on the scalar field as
\eqn\pkag{\eqalign{  \phi(\rho,z) \rt  \phih(\rho,z)   &  =\big(1  -\Lambda(z)\big)  \phi(\rho,z)\cr & =\big(1  -\Lambda(z)\big)\int \! d^2z' G_{b\p}(\rho,z;z_2)\phi_-(z')   }}
The leading asymptotic behavior of the transformed scalar solution is
\eqn\pkah{ \phih(\rho,z) \sim \big( 1-\Lambda(z)\big)  \phi_-(z) e^{-(1-\lambda)\rho}~,\quad \rho \rt \infty  }
from which we read off the relation between the original and transformed sources
\eqn\pkai{  \phih_-(z) = \big( 1-\Lambda(z)\big)  \phi_-(z) }
Inverting to first order in $\Lambda$, taking $\phih_-(z) = \mu_\phi \delta^{(2)}(z-z_2)$ gives us an expression for the original source
\eqn\pkaj{ \phi_-(z) = \mu_\phi \big( 1+ \Lambda(z)\big)\delta^{(2)}(z-z_2) }
We now insert this result in \pkag\ and compute the asymptotic behavior for $z \neq z_{2,3}$.  Retaining just the term of order $\mu_\phi \mu_A$, we find
\eqn\pkak{ \phih(\rho,z)\sim  {\lambda \over \pi} \mu_\phi \left({\Lambda(z_2) \over |z-z_2|^{2(1+\lambda)} } -  {\Lambda(z) \over |z-z_2|^{2(1+\lambda)} } \right) e^{-(1+\lambda)\rho}~,\quad \rho \rt \infty
  }
from which we read off
\eqn\pkal{ \Oc(z_1) = { \lambda  \mu_\phi B_\phi  \over \pi}\Bigg( { \Lambda(z_2) - \Lambda(z_1)\over  |z_{12}|^{2(1+\lambda)} } \Bigg)}
Inserting the formula for $\Lambda$ given in \pkaf\ and using \pkad\ we arrive at
\eqn\pkam{\eqalign{ \langle \Oc(z_1) \Ocb(z_2) J^{(1)}(z_3)\rangle & = { \lambda B_\phi\over 2\pi^2 }\left(z_{12} \over z_{13} z_{23} \right){1 \over  |z_{12} |^{2(1+\lambda)} } \cr
& = {1\over 2\pi} \left(z_{12} \over z_{13} z_{23} \right)  \langle \Oc(z_1) \Ocb(z_2) \rangle              } }

This simple example illustrates the power of our approach. Starting from the solution for a free scalar in AdS$_3$, as given by the bulk to boundary propagator, all we need to do is to perform a gauge transformation to generate a solution with the required asymptotic behavior of the gauge field.  From this solution we read off the scalar vev, and thence the three-point function.

\subsec{General spin correlators}
We now establish an algorithm for computing $\langle\Oc_{\pm}(z_1)\overline{\Oc}_{\pm}(z_2) J^{(s)}(z_3)\rangle$ for arbitrary $s$, where we take $\Oc_{\pm}$ and its complex conjugate to be dual to the complex scalar field $C^1_0$ in either standard (+) or alternate $(-)$ quantization. We will work in the standard quantization throughout and include the alternate quantization at the end by taking $\l\rar-\l$. In addition, the case of the second complex bulk scalar $\widetilde{C}^1_0$, dual to an operator $\widetilde{\Oc}_{\pm}$ and its complex conjugate, will be read off afterwards.

Our starting point is the equation \gaa\ which we reproduce here:
\eqn\gmbaa{dC +A\star C - C\star \Ab=0}
This equation is invariant under the \hsl$\oplus$\hsl\ gauge invariance
\eqn\gmba{\eqalign{&A \rar A + d\Lambda + [A,\Lambda]_{\star}\cr
&\Ab \rar \Ab + d\overline{\Lambda} + [\Ab,\overline{\Lambda}]_{\star}\cr
&C \rar C + C\star \overline{\Lambda}-\Lambda\star C }}

Starting from AdS, we introduce a spin-$s$ source in the unbarred sector by performing a chiral gauge transformation with parameter
\eqn\gma{\eqalign{\Lambda(\rho,z)  &=
\sum_{n=1}^{2s-1} {1 \over  (n-1)! }(-\p)^{n-1} \Ls(z) e^{(s-n)\rho}V^s_{s-n}\cr} }
This generates the desired source term in $A_{\zbar}$,
\eqn\gmc{\eqalign{
\delta A_{\zb} & =  \p_{\zb} \Lambda^{(s)} e^{(s-1)\rho} V^s_{s-1}  +\cdots }}
and a conjugate current in $A_z$,
\eqn\gmb{\eqalign{ \delta A_z & =  {1\over (2s-2)!}\p^{2s-1} \Lambda^{(s)}  e^{-(s-1)\rho} V^s_{-(s-1)}   \cr}}

Isolating the lowest component of the transformed $C$ field using \gmba, which we denote with a hat, gives
\eqn\gmm{\widehat{C}^1_0 =C^1_0 + (\delta C)^1_0 = C^1_0-(\Lambda\star C)^1_0}
Using \gma\ and \gb\ we can compute $(\delta C)^1_0$:
\eqn\gsva{\eqalign{(\delta C)^1_0
&= -\sum_{n=1}^{2s-1}{1\over (n-1)!}(-\p)^{n-1}\Ls\cdot {1\over 2}g^{ss}_{2s-1}(s-n,n-s) C^s_{-(s-n)}e^{(s-n)\rho}}}

This depends on an arbitrary component $C^s_{-(s-n)}$ of the master field $C$ in AdS. As discussed previously, these are all fixed on-shell in terms of $C^1_0$ and its derivatives. Once we write $(\delta C)^1_0$ in terms of the AdS scalar $C^1_0$ to obtain an expression analogous to \pkag, the remaining work follows the spin-1 example of subsection 4.1.\vs

To organize the calculation, we first derive a general formula for the correlator, delaying presentation of the explicit formulae for $C^s_{-(s-n)}$ to the next subsection. We can rewrite \gsva\ as
\eqn\gmma{(\delta C)^1_0 = D^{(s)}C^1_0}
for some $s$-dependent differential operator $\Ds$ which contains derivatives $(\p,\pr)$. Substitution for the $C^s_{-(s-n)}$ will reveal\foot{See equations 4.40, 4.42.} that in the sum \gsva, only the terms for which $n\leq s$ will be needed for our computation: these have no $\rho$-dependence, while the $n>s$ terms decay at the AdS boundary. This will imply that $D^{(s)}$ is of order $s-1$ in derivatives $\p$, so we can decompose $D^{(s)}$ as
\eqn\go{D^{(s)} = \sum_{n=1}^{s}f^{s,n}(\l,\pr)\p^{n-1}\Ls\p^{s-n}}
All of the nontrivial information about the higher spin deformation is hidden in the functions $f^{s,n}(\l,\pr)$.

We now switch to the notation
\eqn\swa{C^1_0 \equiv \phi}
The leading asymptotic behavior of the transformed scalar is
\eqn\gppa{\widehat{\phi}(\rho,z) \sim (1+D^{(s)})e^{-(1-\l)\rho}{\phi}_-(z)~, \quad \rho\rar\infty}
To move the $\Ds$ through the $\rho$-dependent prefactor, we define
\eqn\gq{\eqalign{D^{(s)}_{\pm} &\equiv D^{(s)}(\pr \rar -(1\pm \l))\cr}}
and likewise for $f^{s,n}_{\pm}(\l)$.
Then setting the transformed source  $\widehat{\phi}_-(z)=\mu_{\phi}\delta^{(2)}(z-z_2)$
and inverting to linear order,
\eqn\grs{\phi_-(z) = \mu_{\phi}(1-D^{(s)}_-)\delta^{(2)}(z-z_2)}
The asymptotic behavior of the transformed scalar for $z\neq z_{2}$ now reads, omitting the leading part that is local in the higher spin source,
\eqn\gss{\widehat{\phi}_+(z) = {\l \mu_{\phi}\over \pi}\int d^2z' (1+D^{(s)}_+(z)){(1-D^{(s)}_-(z'))\delta^{(2)}(z'-z_2)\over |z-z'|^{2(1+\l)}}~, \quad \rho\rar\infty}
 
It is now important to know what coordinates derivatives are acting on, so we re-institute the subscript on derivatives: $\p \rar \p_z$. Isolating the piece of order $\mu_{\phi}\Ds$ and placing our scalar operator at the boundary point $z=z_1$, we have
\eqn\gst{\widehat{\phi}_+(z_1) = {\l \mu_{\phi}\over \pi}\left[D^{(s)}_+(z_1)\cdot{1\over |z_{12}|^{2(1+\l)}} - \int d^2z' {D^{(s)}_-(z')\delta^{(2)}(z'-z_2)\over |z_1-z'|^{2(1+\l)}}\right]}
We want to integrate the second piece by parts, which was the point of the definition \go. Writing the integral as
\eqn\gsu{\int d^2z' {D_-(z')\delta^{(2)}(z'-z_2)\over |z_1-z'|^{2(1+\l)}} = \sum_{n=1}^sf^{s,n}_-(\l)\int d^2z' {\p_{z'}^{n-1}a \p_{z'}^{s-n}\delta^{(2)}(z'-z_2)\over |z_1-z'|^{2(1+\l)}}}
integrating by parts and ignoring boundary terms, the $n$'th term is
\eqn\gsua{\eqalign{&(-1)^{s-n}f^{s,n}_-(\l)\p_{z_2}^{s-n}\left({\p_{z_2}^{n-1}\Ls \over |z_{12}|^{2(1+\l)}}\right)\cr
= &(-1)^{s-n}f^{s,n}_-(\l)\sum_{j=0}^{s-n}{s-n\choose j}[\p_{z_2}^{n-1+j}\Ls] \p_{z_2}^{s-n-j}{1 \over |z_{12}|^{2(1+\l)}}  }}
Making use of the identity,
\eqn\gsub{\eqalign{\p_{z'}^n {1\over |z_1-z'|^{2(1+\l)}} &= (-1)^n \p_{z_1}^n {1\over |z_1-z'|^{2(1+\l)}} = {\Gamma(\l+n+1)\over\Gamma(\l+1)}{1\over (z_1-z')^n}{1\over|z_1-z'|^{2(1+\l)}}\cr
}}
 one can reduce \gst\ to the following expression in terms of the $f^{s,n}_{\pm}(\l)$ and the transformation parameter $\Ls$:
\eqn\gsv{\eqalign{\widehat{\phi}_+(z_1) &= {\l \mu_{\phi}\over \pi |z_{12}|^{2(1+\l)}}\Bigg[\sum_{n=1}^s\left({-1\over z_{12}}\right)^{s-n}\Big\lbrace f_+^{s,n}(\l){\Gamma(\l+s-n+1)\over\Gamma(\l+1)}\p_{z_1}^{n-1}\Ls\cr &-f_-^{s,n}(\l)\sum_{j=0}^{s-n}{s-n\choose j}{\Gamma(\l+s-n-j+1)\over\Gamma(\l+1)}(\p_{z_2}^{n-1+j}\Ls)z_{12}^{j}\Big\rbrace\Bigg]\cr}}

\gsv\ is the general expression for the scalar vev in the presence of a higher spin perturbation, to first order in each of the scalar and higher spin sources. For a higher spin delta function source at $z_3$, we take (cf. \gmc)
\eqn\gva{\Ls = {1\over 2\pi}{1\over z-z_3}}
and \gsv\ becomes
\eqn\gvb{\eqalign{\widehat{\phi}_+(z_1) &= {(-1)^{s-1}\l \mu_{\phi}\over 2\pi^2 |z_{12}|^{2(1+\l)}}\sum_{n=1}^s{1\over z_{12}^{s-n}} \Bigg\lbrace f_+^{s,n}(\l){\Gamma(\l+s-n+1)\over\Gamma(\l+1)}(n-1)! {1\over z_{13}^n}\cr &-f_-^{s,n}(\l){1\over z_{23}^n}\sum_{j=0}^{s-n}(-1)^j{s-n\choose j}{\Gamma(\l+s-n-j+1)\over\Gamma(\l+1)}(n-1+j)! \left({z_{12}\over z_{23}}\right)^j\Bigg\rbrace\cr}}

Therefore, reading off the three-point function $\langle\Oc_+(z_1)\overline{\Oc}_+(z_2) J^{(s)}(z_3)\rangle$ boils down to knowing the functions $f^{s,n}_{\pm}(\l)$ encoding the change in the scalar under gauge transformation.

Before turning to the problem of determining these functions $f^{s,n}_{\pm}(\l)$, we note that a conformally symmetric result has the property
\eqn\gvd{\langle\Oc(z_1)\overline{\Oc}(z_2) J^{(s)}(z_3)\rangle = (-1)^s\langle\Oc(z_2)\overline{\Oc}(z_1) J^{(s)}(z_3)\rangle}
This property is not manifest in our formula \gvb, but it implies that, resorting to the notation of \gst, the correct solution is given by
\eqn\gve{\widehat{\phi}_+(z_1) = {\l \mu_{\phi}\over \pi}\left[D^{(s)}_+(z_1)+(-1)^sD^{(s)}_+(z_2)\right]\cdot{1\over |z_{12}|^{2(1+\l)}}}
In terms of the $\fsn$ this looks like
\eqn\gvf{\eqalign{\widehat{\phi}_+(z_1) &= {(-1)^{s-1}\l \mu_{\phi}\over 2\pi^2 |z_{12}|^{2(1+\l)}}\sum_{n=1}^s{f_+^{s,n}(\l)\over z_{12}^{s-n}} {\Gamma(\l+s-n+1)\over\Gamma(\l+1)}(n-1)! \left({1\over z_{13}^n}+{(-1)^n\over z_{23}^n}\right)\cr &\cr}}
where the extra $(-1)^n$ comes from sign changes under derivatives acting on ${1\over |z_{12}|}$ (cf. \gsub). One can show, by using \gvb\ and \gvf\ and equating terms with the same number of powers of $z_{12}$, that this in turn implies the following unobvious relations:

\eqn\gvg{f_+^{s,j}(\l) = -\sum_{n=1}^s(-1)^nf_-^{s,n}(\l){s-n\choose j-n}}
for some fixed $j$.

Upon solving for the $\fsn$ generated by the gauge transformation \gma, we will show that \gvg\ is indeed satisfied.
\subsec{The AdS master field $C$}
To write the $\fsn$ as defined by \gsva, \gmma\ and \go,
we need a formula for all components of the master field $C$ in AdS, written in terms of $C^1_0\equiv\phi$. This amounts to solving \gi. We first solve for the minimal components $C^{m+1}_{\pm m}$ in terms of $\phi$, then for the non-minimal components $C^{s\neq m+1}_{\pm m}$ in terms of the $C^{m+1}_{\pm m}$, and finally we put the two together. \vs

\noindent
{\bf Minimal components $C^{m+1}_{\pm m}$:}
\vskip.1cm

\noindent  Taking $m\rar-m$ and $s=m+1$ in the second equation in \gi\ gives
\eqn\gp{\eqalign{\pz C^{m+1}_{-m} + {\ep\over 2} g^{2(m+2)}_{3}(1,-m-1)C^{m+2}_{-m-1}=0}}
where we recall that for some component $C^s_n$ one needs $|n|\leq s-1$.
Solving recursively yields the following expression:
\eqn\gq{C^{m+1}_{-m}=\left(\prod_{p=2}^{m+1}g^{2p}_3(1,1-p)\right)^{-1}(-2\emp\pz)^{m} \phi}
A similar analysis along $\zbar$ yields
\eqn\gr{C^{m+1}_{m}=\left(\prod_{p=2}^{m+1}g^{2p}_3(-1,p-1)\right)^{-1}(2\emp\pzb)^{m} \phi}
where we have used the fact that $g^{st}_3(m,n)=g^{ts}_3(n,m)$.
\vs

\noindent
{\bf Non-minimal components
$C^{s\neq m+1}_{\pm m}$:}
\vskip.1cm

\noindent
From the $\rho$ equations in \gi, one can see that, say, the component $C^2_0$ will have the same structure, when written in terms of $\phi$, as $C^3_1$ in terms of $C^2_1$, and so on. In general, components with fixed $s-|m|$ when expressed as a function of their respective minimal components will have the same form in terms of structure constants.

The solution is
\eqn\gt{\eqalign{C^s_{\pm m} &= (-1)^{s-1-m}\left(\prod_{p=2+m}^sg^{p2}_3(m,0)\right)^{-1}\left[ \sum_{\a=0}^{\lfloor{s-1-m\over 2}\rfloor}A_{\a}(s,m) \pr^{s-2\a-m-1} \right] C^{m+1}_{\pm m} \cr}}
The $A_{\a}(s,m)$ are defined as
\eqn\gu{\eqalign{&A_{\a}(s,m) = (-2)^{\a}\sum_{i_1\ldots i_{\a}}\prod_{k=1}^{\a}g^{i_k2}_3(m,0)\cr}}
with indices subject to
\eqn\gv{\eqalign{&2k+m\leq i_k \leq 2k+s-1-2\a\cr
&i_k \geq i_{k-1}+2~, \quad \forall ~k\geq 2}}
In appendix D, we present expressions for several of the $C^s_0$ obtained from solving \gi\ directly to facilitate easy comparison with \gt.\vs

Via \gq, \gr\ and \gt, one has all components of $C$ in terms of the fundamental scalar field, $\phi$. These formulae also justify our previous statements about subleading terms, and about $\Ds$ being order $s-1$ in derivatives $\p$.

With these results in hand, we combine \gq\ and \gt\ at $m=s-n$ to write $C^s_{-(s-n)}$ in terms of $\phi$, and plug into \gsva. The result can be written
\eqn\gsx{\eqalign{(\delta C)^1_0 =&\sum_{n=1}^{s}f^{s,n}_{\pm}(\l)\pz^{n-1}\Ls\pz^{s-n} \phi~ +({\rm  subleading})\cr = &\sum_{n=1}^s(-1)^s{2^{s-n-1}\over (n-1)!} {\cal{F}}_{\pm}(s,s-n;\lambda)\pz^{n-1}\Ls \pz^{s-n}\phi}}
where
\eqn\gszb{\eqalign{{\cal{F}}_{\pm}(s,s-n;\lambda)
&=g^{ss}_{2s-1}(s-n,n-s)\cr&\times \left(\prod_{p=2+s-n}^sg^{p2}_3(s-n,0)\right)^{-1}\left(\prod_{p=2}^{s-n+1}g^{p2}_3(1-p,1)\right)^{-1} \cr
&\times\left[ \sum_{\a=0}^{\lfloor{n-1\over 2}\rfloor}A_{\a}(s,s-n)(\pm\l+s-n+1)^{n-1-2\a}
\right] \cr}}
with $A_{\a}(s,s-n)$ defined as in \gu\ and \gv. We have dropped the subleading terms in the second line of \gsx. To obtain this result we have used the replacement $\pr \rar -(1\pm\l)$ and the identity
\eqn\gsgh{\sum_{\chi=0}^{n-1-2\a}{n-1-2\a\choose\chi}(s-n)^{n-1-2\a-\chi}(1\pm\l)^{\chi} = (\pm\l+s-n+1)^{n-1-2\a}}
From \gsx, we easily read off the reduced expression for the functions $f^{s,n}_{\pm}(\l)$ :
\eqn\gsza{f^{s,n}_{\pm}(\l) = (-1)^s{2^{s-n-1}\over (n-1)!} {\cal{F}}_{\pm}(s,s-n;\lambda)}

We can now plug this into our general formula \gvb\ and read off the correlator. If we are to obtain a conformally invariant result, it must also satisfy the conformal identity \gvg, in which case formula \gvf\ is equally valid.

Of course, \gsza\ is a rather complicated function when expressed in terms of structure constants, and combined with \gvf\ it is not at all clear that we will arrive at our desired result. Accordingly we expect dramatic simplification of the $\fsn$ when the structure constants are written out explicitly.
\subsec{Final result}
We set out to compute the $f^{s,n}_{\pm}(\l)$ for low values of $n$, using the formulae \gszb, \gsza. The results up to $s=8$ are:

\eqn\fsdd{\eqalign{f^{s,1}_{\pm}(\l)&= {(-1)^s}\cr
f^{s,2}_{\pm}(\l)&= {(-1)^s\over 2}{\Gamma(s\pm\l)\over\Gamma(s-1\pm\l)}\cr
f^{s,3}_{\pm}(\l)&= {(-1)^s\over 4}{s-2\over 2s-3}{\Gamma(s\pm\l)\over\Gamma(s-2\pm\l)}\cr
f^{s,4}_{\pm}(\l)&= {(-1)^s\over 24}{s-3\over 2s-3}{\Gamma(s\pm\l)\over\Gamma(s-3\pm\l)}\cr
f^{s,5}_{\pm}(\l)&= {(-1)^s\over 96}{(s-3)(s-4)\over (2s-3)(2s-5)}{\Gamma(s\pm\l)\over\Gamma(s-4\pm\l)}\cr
f^{s,6}_{\pm}(\l)&= {(-1)^s\over 960}{(s-4)(s-5)\over (2s-3)(2s-5)}{\Gamma(s\pm\l)\over\Gamma(s-5\pm\l)}\cr
f^{s,7}_{\pm}(\l)&= {(-1)^s\over 5760}{(s-4)(s-5)(s-6)\over (2s-3)(2s-5)(2s-7)}{\Gamma(s\pm\l)\over\Gamma(s-6\pm\l)}\cr
f^{s,8}_{\pm}(\l)&= {(-1)^s\over 80640}{(s-5)(s-6)(s-7)\over (2s-3)(2s-5)(2s-7)}{\Gamma(s\pm\l)\over\Gamma(s-7\pm\l)}\cr
}}
Indeed, these are compact expressions. By induction, we obtain a tidy formula for these functions:
\eqn\gvk{f^{s,n}_{\pm}(\l) = (-1)^s{\Gamma(s\pm\l)\over\Gamma(s-n+1\pm\l)}{1  \over 2^{n-1}(2\lfloor{n\over 2}\rfloor-1)!!\lfloor{n-1\over 2}\rfloor!}\prod_{j=1}^{\lfloor{n- 1\over 2}\rfloor}{s+j-n\over 2s-2j-1}}

Encouragingly, one can check spin-by-spin that this expression indeed satisfies the conformal identity \gvg. Plugging this into \gvf\ and including the alternate quantization  $\l \rar -\l$ then yields the final answer for the correlator:
\eqn\gvl{\eqalign{\langle{\Oc}_{\pm}(z_1)\overline{\Oc}_{\pm}(z_2) J^{(s)}(z_3)\rangle &= {(-1)^{s-1}B_{\phi}\over 2\pi^2 |z_{12}|^{2(1\pm\l)}}{\Gamma(s)^2\over\Gamma(2s-1)}{\Gamma(s\pm\l)\over\Gamma(\pm\l)}\left({z_{12}\over z_{13}z_{23}}\right)^s\cr &= {(-1)^{s-1}\over 2\pi }{\Gamma(s)^2\over\Gamma(2s-1)}{\Gamma(s\pm\l)\over\Gamma(1\pm\l)}\left({z_{12}\over z_{13}z_{23}}\right)^s\langle{\Oc}_{\pm}(z_1)\overline{\Oc}_{\pm}(z_2)\rangle \cr}}
This is manifestly conformally invariant, and the overall coefficient is the main result.

While we lack a rigorous proof of \gvk\ and \gvl, we have confirmed \gvk\ up to $n=13$; assuming \gvk, it is simple to confirm \gvl\ at any desired spin. \vs

Recalling the discussion at the end of section 2, there exists a second projection of the master field $C$ that gives rise to the equation
\eqn\galt{d\widetilde{C} + \Ab\star \widetilde{C} - \widetilde{C}\star A=0}
The lowest component of $\widetilde{C}$ is also a complex scalar field, dual to a scalar CFT  operator $\widetilde{\Oc}$ and its complex conjugate. This equation is simply related to \gmbaa\ by the exchange $A \leftrightarrow \Ab$, which flips the sign of all odd spin tensors. Therefore, the result for the tilded correlator is simply our result \gvl\ with a $(-1)^s$ removed:
\eqn\gvla{\langle\widetilde{\Oc}_{\pm}(z_1)\overline{\widetilde{\Oc}}_{\pm}(z_2) J^{(s)}(z_3)\rangle = -{1\over 2\pi }{\Gamma(s)^2\over\Gamma(2s-1)}{\Gamma(s\pm\l)\over\Gamma(1\pm\l)}\left({z_{12}\over z_{13}z_{23}}\right)^s\langle\widetilde{\Oc}_{\pm}(z_1)\overline{\widetilde{\Oc}}_{\pm}(z_2)\rangle }\vs
In the context of the $W_N$ coset CFT duality conjecture, one scalar is in standard quantization and the other in alternate quantization. The operation of flipping the sign of odd spin fields corresponds to a charge conjugation operation in the CFT.

These results match and extend the bulk calculations of \ChangMZ\ which were restricted to $\l=1/2$. To further compare to the CFT, we now compute the same correlators using CFT considerations, again for all $\l$ and $s$, and find perfect agreement with the bulk.

\newsec{Three-point correlators from  CFT}

We now shift focus and consider the constraints on three-point functions due to the existence of a higher-spin current algebra.  Our considerations will be entirely based on symmetry, and in particular on the existence of $\Wc_\infty[\lambda]$ current algebra.  If these currents are not present in the CFT, then even before considering scalar operators there will be a mismatch between bulk and boundary correlators involving only currents.  So we will assume the existence of this symmetry algebra, and then see what constraints this imposes on the scalar-scalar-current three-point functions.  Our computations in this section are along the same lines as in \GaberdielWB, but generalized to arbitrary $s$.

Suppose we have a spin-s current $J^{(s)}(z)$, and scalar primary operator $\Oc(z,\zb)$,\foot{We use the shorthand $\Oc(z,\zb)=\Oc(z)$ in what follows.} whose OPE has the following leading singularity,
\eqn\pkba{ J^{(s)}(z) \Oc(0) \sim {A(s) \over z^s} \Oc(0) + \cdots }
The three-point function is then related to the scalar two-point function as
\eqn\pkbb{ \langle \Oc(z_1) \Ocb(z_2) J^{(s)}(z_3)\rangle = A(s) \left( z_{12} \over z_{13}z_{23}\right)^s \langle \Oc(z_1) \Ocb(z_2) \rangle  }
Writing the mode expansion
\eqn\pkbc{ J^{(s)}(z) =-{1\over 2\pi}  \sum_{m=-\infty}^\infty {W_m^{(s)}\over z^{m+s} } }
the zero modes act on the primary state $|\Oc\rangle$ as
\eqn\pkbd{ J^{(s)}_0 |\Oc\rangle = -2\pi A(s) |\Oc\rangle  }

In the case of interest, the currents $J^{(s)}(z)$, $s = 2,3, \ldots$, obey the $\Wc_\infty[\lambda]$ current algebra.  The wedge modes are defined to be those that annihilate the vacuum state, and are given by
\eqn\pkbe{ V^s_m = J^{(s)}_m~,\quad   |m| < s }
In general, these wedge modes do not yield  a closed subalgebra of  $\Wc_\infty[\lambda]$, as their commutators yield modes outside the wedge; in particular, this is due to the nonlinearities present in $\Wc_\infty[\lambda]$.  However, the nonlinear terms are suppressed at large central charge, and so in the limit $c\rt \infty$ the wedge modes do define a subalgebra of $\Wc_\infty[\lambda]$  -- the wedge subalgebra.   This subalgebra is \hsl.  To exploit this simplification, for the remainder of this section we will assume that we are working in the limit of large central charge.  See \GaberdielWB\ for further discussion.

Furthermore, these considerations fix the relative normalization of the bulk and boundary currents.  In particular, we have defined our conventions such that the currents \pkbc\ are equal to the currents derived from the bulk.

Acting on $|\Oc\rangle$ with the wedge modes, we obtain a representation of \hsl, and we can  therefore use the representation theory of \hsl\ to determine the zero mode eigenvalues appearing in \pkbd, and thence the three point function \pkbb.    The Virasoro zero mode eigenvalue is fixed by the scaling dimension of $\Oc$,
\eqn\pkbf{ V^2_0 |\Oc_\pm  \rangle  = {1\over 2}(1\pm \lambda) |\Oc_\pm \rangle}
where we've now introduced the pair of scalar operators $\Oc_\pm$.
We need to compute the remaining eigenvalues.

To proceed, it is useful to build up the \hsl\ generators in terms of SL(2) generators.   We write  $V^2_1 = L_1,  V^2_0= L_0, V^2_{-1}=L_{-1}$, which obey the SL(2) algebra
\eqn\pkbg{ [L_1,L_{-1}]=2L_0~,\quad [L_0,L_1] = -L_1~,\quad [L_0,L_{-1}]=L_{-1}}
We then construct the $V^s_m$ generators as\foot{As will be discussed at the end of this section, we obtain a second inequivalent representation by appending a factor of $(-1)^s$ to the generators.}
\eqn\pkbi{ V^s_m = (-1)^{m-1} {(m+s-1)!\over (2s-2)!}\underbrace{ \big[ L_{-1}, \ldots [L_{-1}, [L_{-1},}_{s-1-m} L_1^{s-1}]] \big]}
where in addition we mod out by the ideal obtained by fixing the SL(2) quadratic Casimir as
\eqn\pkbh{ C_2 = L_0^2 -{1\over 2}(L_1 L_{-1} +L_{-1}L_1) = {1\over 4}(\lambda^2-1)}
We can use this to work out the zero mode eigenvalues, as illustrated by the first nontrivial case:
\eqn\pkbj{\eqalign{ s=3:\quad &V^3_0|\Oc_\pm \rangle =- {1\over 12} \big[ L_{-1},[L_{-1},L_1^2] \big] =\left( {1\over 3} C_2 -L_0^2 \right)|\Oc_\pm\rangle= -{1\over 6}(\lambda\pm2)(\lambda\pm1)|\Oc_\pm \rangle
    }}
In the last step we used  \pkbf\ and  \pkbi.  Working out further examples by brute force quickly gets tedious, so we adopt a more indirect approach.

First, it's easy to see that $V^s_0$ will be a polynomial in $\lambda$ of degree $s-1$, as illustrated in \pkbj.  This follows, since the terms in \pkbi\ obtained after working out all the commutators will each have $s-1$ generators, and using either \pkbh\ or $L_0 = {1\over 2}(1\pm \lambda)$ will convert a generator into at most one power of $\lambda$.

Next consider taking $\lambda =N$, a positive integer.   In this case, after factoring out an ideal, \hsl\ becomes SL(N), which we can represent in terms of $N\times N$ matrices.   In this representation the generators $V^s_m$ with $s>N$ all vanish identically when they are constructed using \pkbi, and in particular this holds for the zero modes $V^s_0$.  We also note that the eigenvalues of $L_0 = V^2_0$ in the $N\times N$  matrix representation are: $-{N-1\over 2}, -{N-1\over 2}+1, \ldots , {N-1\over 2} $. Note that the smallest eigenvalue coincides with ${1\over 2}(1-\lambda)$, i.e. with the eigenvalue of $L_0$ acting on $|\Oc_-\rangle$.    Together, these facts imply that
$V^s_0 |\Oc_-\rangle = 0$ for $\lambda = 1, 2, \ldots, s-1$.  Combining this with the statement in the previous paragraph, we fix the $\lambda$ dependence of the zero-mode eigenvalues to be
\eqn\pkbk{\eqalign{ V^s_0 |\Oc_-\rangle &= N(s) [\lambda - (s-1) ] \cdots [\lambda-2][\lambda -1]|\Oc_-\rangle \cr
& = N(s)(-1)^{s-1} {\Gamma(s-\lambda) \over \Gamma(1-\lambda)}|\Oc_-\rangle  }}
for some prefactor $N(s)$.
To obtain the eigenvalues for $|\Oc_+\rangle$ we simply flip the sign of $\lambda$, and we arrive at
\eqn\pkbl{  V^s_0 |\Oc_\pm \rangle = N(s)(-1)^{s-1} {\Gamma(s\pm \lambda) \over \Gamma(1\pm \lambda)}|\Oc_\pm \rangle  }

To fix $N(s)$ we can pick some convenient value of $\lambda$.  If we take $\lambda =1/2$ then we can represent SL(2) as
\eqn\pkbm{ L_0 = -{1\over 4} (x\p_x + \p_x x)   ~,\quad L_{-1} = {1\over 2}  \p_x^2~,\quad L_1  = {1\over 2}  x^2 }
We can then construct hs$[{1\over 2}]$ using \pkbi.

Now, at $\lambda =1/2$ we have $L_0 |\Oc_+\rangle  = {3\over 4} |\Oc_+\rangle$, which in the representation \pkbm\ is achieved by taking $L_0$ to act on the function $x^{-2}$.   As shown in appendix C the eigenvalues of the other zero mode generators are then computed to be
\eqn\pkbn{ V^s_0 x^{-2} =  {(-1)^s[(s-1)!]^2 (2s-1)!!  \over 2^{s-1}(2s-2)!} x^{-2}  }
Comparing this with the $\Oc_+$ version of \pkbl\  at $\lambda =1/2$ yields
\eqn\pkbo{N(s) =  - {\Gamma(s)^2 \over \Gamma(2s-1)} }
which gives us our final result for the zero mode eigenvalues

\eqn\pkbp{  V^s_0 |\Oc_\pm \rangle = (-1)^s{ \Gamma(s)^2 \over \Gamma(2s-1)}{\Gamma(  s\pm \lambda) \over \Gamma(1\pm \lambda) }|\Oc_\pm \rangle  }

Using \pkba\ and \pkbd, we have therefore fixed the three-point function to be
\eqn\pkbq{ \langle \Oc_\pm (z_1) \Ocb_\pm (z_2) J^{(s)}(z_3)\rangle = {(-1)^{s-1}\over 2\pi}  { \Gamma(s)^2 \over \Gamma(2s-1)}{\Gamma(  s\pm \lambda) \over \Gamma(1\pm \lambda) } \left( z_{12} \over z_{13}z_{23}\right)^s \langle \Oc_\pm (z_1) \Ocb_\pm (z_2) \rangle  }
This agrees perfectly with the correlator obtained from the bulk, \gvl.

Now let us come back to  the footnote accompanying \pkbi.  \hsl\ admits the automorphism $V^s_m \rt (-1)^s V^s_m$, which can be thought of as charge conjugation.   If we had instead used the charge conjugate generators in \pkbi\ then a factor of $(-1)^s$ would have propagated through to the final result \pkbq.  Equivalently, starting from $\Oc_\pm$ we can consider operators $\widetilde{\Oc}_\pm$ that transform as their charge conjugates.  The three-point function of such operators is therefore
\eqn\pkbr{ \langle \widetilde{\Oc}_\pm (z_1) \overline{\widetilde{\Oc}}_\pm (z_2) J^{(s)}(z_3)\rangle = -{1\over 2\pi}  { \Gamma(s)^2 \over \Gamma(2s-1)}{\Gamma(  s\pm \lambda) \over \Gamma(1\pm \lambda) } \left( z_{12} \over z_{13}z_{23}\right)^s \langle \widetilde{\Oc}_\pm (z_1) \overline{\widetilde{\Oc}}_\pm (z_2)  \rangle  }
This agrees with the bulk result \gvla.

\newsec{Discussion}

\subsec{Comparison with previous results}

The results \pkbq-\pkbr\ derived from CFT considerations agree perfectly with the correlators derived from the higher spin theory in the bulk, namely \gvl\ and \gvla.  Before discussing the implications of this agreement, let us compare to previous work.
To this end,  we note that in our normalization the current-current two-point function is
\eqn\pkda{ \langle J^{(s)}(z) J^{(s)}(0)\rangle = {3k \over 2^{2s-1} \pi^{5/2} } {\sin (\pi \lambda) \over \lambda (1-\lambda^2)}{\Gamma(s)\Gamma(s-\lambda)\Gamma(s+\lambda) \over \Gamma(s-{1\over 2}) }{1\over z^{2s}} }
The derivation of this result proceeds in the same fashion as led to \chha.  This normalization is the ``natural" one, since the wedge modes  defined in \pkbc\ then obey the \hsl\ algebra with standard normalization.

In \ChangMZ\ three-point correlators were computed from the bulk for arbitrary $s$ and $\lambda=1/2$; and in the 't Hooft limit of the $W_N$ minimal model for $s=3$ and arbitrary $\lambda$.   Their currents are normalized to
$\langle J^{(s)}(z)J^{(s)}(0)\rangle = z^{-2s}$.   Under the identification $\Oc_+^{{\rm here}} = \overline{\Oc}_+^{\rm CY}$ and $\widetilde{\Oc}_-^{{\rm here}} = \overline{\Oc}_-^{\rm CY}$, and taking into account the different normalizations for the currents, we verify that our results reduce to those of \ChangMZ\ for the special values of $s$ and $\lambda$.

In \AhnBY\ the three-point function for $s=4$ was computed in the 't Hooft limit of the $W_N$ minimal model. The normalization of the current was not specified, but a normalization independent ratio was obtained (see equation 3.35 therein).  The corresponding ratio obtained from our result is
\eqn\pkdb{ { \langle \Oc_+(z_1) \overline{\Oc}_+(z_2) J^{(4)}(z_3) \rangle \over \langle \widetilde{\Oc}_-(z_1) \overline{\widetilde{\Oc}}_-(z_2) J^{(4)}(z_3) \rangle} = { (1+\lambda)(2+\lambda)(3+\lambda)\over(1-\lambda)(2-\lambda)(3-\lambda)}}
which agrees with \AhnBY.

\subsec{Comments}

We now discuss the implications of our agreement between bulk and boundary.   On the CFT side, what went into the computation was the assumption that the CFT has a symmetry algebra containing \hsl, along with scalar operators of the correct dimension; everything else followed from \hsl\ representation theory.   So any CFT with these properties will have three-point functions that match those of the bulk.  One way that \hsl\ symmetry can emerge is if the CFT has $\Wc_\infty[\lambda]$ symmetry, and the central charge is taken to infinity. Then \hsl\ is identified with the wedge subalgebra of $\Wc_\infty[\lambda]$.

Now consider the case of the $\Wc_N$ minimal models proposed by Gaberdiel and Gopakumar as CFT duals of the bulk higher spin theory.  As we have stressed, even before considering the scalars, it is necessary that in the 't Hooft limit the CFT acquire $\Wc_\infty[\lambda]$ symmetry if it is to have a chance of matching with the bulk.   The bulk theory has such a symmetry, and this fixes the form of all correlation functions on the plane involving just currents.  These will not match with the CFT unless the latter also has $\Wc_\infty[\lambda]$ symmetry.

Assuming that the CFT does indeed exhibit \hsl\ symmetry, let's consider the scalars.    The results of this paper establish that if the CFT has scalar operators of the correct dimension, $\Delta = 1\pm \lambda$, then the scalar-scalar-current three-point functions on the plane will match between the bulk and boundary.

As an illustration of these comments, we now establish that a theory of free bosons has correlators that match those of the corresponding bulk theory.  Here by ``correlators" we mean those discussed above: namely pure current correlators, and  scalar-scalar-current correlators, all evaluated on the plane.  Consider the following currents
\eqn\pkca{ J^{(s+2)} = -{1\over 2\pi} {2^{-s-1} (s+2)! \over (2s+1)!!} \sum_{k=0}^s (-1)^k {1\over s+1} \left(\matrix{s+1 \cr k }\right)\left(\matrix{s+1 \cr k+1} \right) \p^{s-k+1} \phib \p^{k+1} \phi  }
where $\phi$ is a complex free boson.  These currents yield the linear algebra $\Wc_\infty^{\rm PRS}$ \prs\ at $c=2$ \refs{\BakasRY,\BergshoeffYD};  the generalization to higher $c$ is obtained by introducing additional copies of the free boson. After a nonlinear redefinition of the currents, $\Wc_\infty^{\rm PRS}$ becomes equivalent to $\Wc_\infty[1]$ \refs{\FigueroaOCV,\GaberdielWB}. For $\lambda=1$ the bulk scalar is dual to  a CFT operator of dimension $2$, and this is
$\Oc = \p \phi \overline{\p} \phib$.  The results of this paper show that the scalar-scalar-current three-point functions of this free boson theory will match those of the higher spin theory in the bulk at $\lambda =1$.  For instance,  it is simple to check this explicitly for the case of the spin-3 current.

A related situation occurs with complex free fermions at $\lambda=0$.  The following currents \BergshoeffYD
\eqn\pkcb{  J^{(s+2)} = {1\over 2\pi} {2^{-s-1} (s+1)! \over (2s+1)!!} \sum_{k=0}^{s+1} (-1)^k \left(\matrix{s+1 \cr k}\right)^2 \p^{s-k+1}\psib \p^k \psi }
realize the algebra $\Wc_{1+\infty}$ \PopeKC\ at $c=1$.  Although $\Wc_{1+\infty}$ is not equivalent to $\Wc_\infty[0]$ due to the presence of the spin-1 current in $\Wc_{1+\infty}$, the wedge subalgebra of $\Wc_{1+\infty}$ yields hs$[0]$; the spin-1 zero mode just yields an operator that commutes with all the other wedge modes.
The scalar operator in this theory is $\Oc = \psib \psi$.   As we have discussed, this is enough structure to guarantee that the scalar-scalar-current correlators (for spin greater than 1) will match those of the bulk theory at $\lambda =0$, and verifying this is straightforward.

Note that we are not making any claims here about a full duality between these free boson/fermion theories and their bulk counterparts. Indeed, without further ingredients it seems clear that the theories cannot be equivalent:  if we simply add $N$ copies of the free fields the CFT will have a $U(N)$ symmetry along with various nonsinglet operators, none of which appear to be present in the bulk, at least classically.

\vskip .5cm
\noindent
{ \bf Acknowledgments}

\vskip .2cm
\noindent
This work was supported in part by the National Science Foundation under Grant No. NSF PHY-07-57702. We are grateful to  M. Gutperle for useful conversations, and the KITP for hospitality. This research was supported in part by the National Science Foundation under Grant No. NSF PHY05-51164. E.P. is supported in part by a UCLA Graduate Division Dissertation Year Fellowship.

\appendix{A}{Lightning review of 3D higher spin gravity coupled to scalars}
In this appendix we present all details necessary for the bulk theory's construction, following \ProkushkinBQ. In particular we derive the equations \intl\ and \intnb\ in section 2.

According to \ProkushkinBQ, the full non-linear system of equations governing the interaction of matter with higher spin gauge fields is formulated in terms of the following generating functions: a spacetime 1-form $W = W_\nu dx^\nu$ as well as spacetime 0-forms $B$ and $S_\alpha.$ The generating functions $W, B$ and $S_\alpha$ depend on spacetime coordinates $x,$ on auxiliary bosonic twistor variables $z_\alpha$ and $y_\alpha$ ($\alpha=1,2$) as well as on two pairs of Clifford elements, $\psi_{1,2}$ and $k,\rho:$
\eqn\eqIIa{ \{ \psi_i, \psi_j \} = 2 \delta_{ij}~, ~~ k\rho = - \rho k~, ~~  k^2 = \rho^2 =1}

Moreover, $\psi_{1,2}$ commute with all other auxiliary variables, and $k,\rho$ obey
\eqn\eqIIab{\eqalign{k y_\alpha &= - y_\alpha k, \quad k z_\alpha = - z_\alpha k \cr
\rho y_\alpha &=y_\alpha \rho, \quad\ \ \rho z_\alpha = z_\alpha \rho}}
Indices on $z_\alpha$ and $y_\alpha$ are raised by $\epsilon^{\alpha\beta}$ and lowered by the rank two antisymmetric tensor $\epsilon_{\beta\alpha}$,
\eqn\eqIIg{ z^\alpha = \epsilon^{\alpha\beta} z_\beta \, , \qquad\quad z_{\alpha} = z^{\beta} \epsilon_{\beta\alpha}}
with $\epsilon^{\alpha\beta} \epsilon_{\beta\gamma} = - \delta^{\alpha}_{\gamma}.$ We follow the convention $\epsilon^{12}=\epsilon_{12}=1.$

Using these properties of the auxiliary variabls the basic fields $W_\nu, B$ and $S_\alpha$ can be expanded in the form
\eqn\eqIId{ A(z,y;\psi_{1,2}, k, \rho | x) = \sum\limits_{b,c,d,e=0}^1 \sum\limits_{m,n=0}^\infty {1\over{m! n!}} A^{\alpha_1\dots\alpha_m \beta_1 \dots \beta_n}_{bcde}(x) \ k^b \rho^c \psi_1^d \psi_2^e \ z_{\alpha_1} \dots z_{\alpha_m} y_{\beta_1} \dots y_{\beta_n}}
where $A$ is either $W_\nu, B$ or $S_\alpha.$ The expression $A^{\alpha_1\dots\alpha_m \beta_1 \dots \beta_n}_{bcde}(x)$ in equation \eqIId\ is an ordinary spacetime function. Note that $A^{\alpha_1\dots\alpha_m \beta_1 \dots \beta_n}_{bcde}(x)$ can be choosen to be symmetric in the indices $(\alpha_1\dots\alpha_m)$ and in the indices $(\beta_1 \dots \beta_n).$

In order to formulate the equations of motion we use the Moyal $\star$-product which acts on the twistors $y$ and $z$ in the following way
\eqn\eqIIh{ f(z,y) \star g(z,y) = {1\over{(2\pi)^2}} \int d^2u \int d^2v \ e^{i (uv)} f(z+u,y+u) g(z-v, y+v)}
where $uv$ is a short-hand notation, $uv = u_\alpha v^\alpha.$
We can verify the following commutation relations:
\eqn\eqIIi{ [y_\alpha, y_\beta]_\star = - [z_\alpha, z_\beta]_\star = 2 i \epsilon_{\alpha\beta}, \qquad [y_\alpha, z_\beta]_\star =0}
where $[a,b]_\star \equiv a \star b - b\star a$ is the commutator with respect to the $\star$-product.

In terms of the generating functions $W= W_\nu dx^\nu, B, S_\alpha$ we are now ready to write down the full non-linear equations of motion \ProkushkinBQ:
\eqn\eqIIe{\eqalign{
dW &= W  \wedge \star W \cr
dB &= W \star B - B \star W \cr
dS_\alpha &= W \star S_\alpha - S_\alpha \star W \cr
S_\alpha \star S^\alpha &= -2i (1+ B \star K) \cr
S_\alpha \star B &= B \star S_\alpha
}}
Here, $K$ -- the so-called Kleinian -- is given by
\eqn\eqIIf{K = k e^{i (z y)}}
These equations \eqIIe\ are invariant under the infinitesimal higher spin gauge transformation
\eqn\eqIIka{\eqalign{ \delta W &= d\epsilon + \epsilon \star W- W \star \epsilon  \cr
\delta B &= \epsilon \star B - B \star \epsilon \cr
\delta S_\alpha &= \epsilon \star S_\alpha - S_\alpha \star \epsilon}}
where $\epsilon$ is the infinitesimal gauge parameter which does not depend on $\rho,$ i.e.
\eqn\eqIIkb{ \epsilon = \epsilon(z,y; \psi_{1,2},k | x)}

We will see that $W$ is the generating function for higher spin gauge fields whereas $B$ is the generating function for the matter fields. $S_{\alpha}$ will describe auxiliary  degrees of freedom.
 
Since the equations of the motion \eqIIe\ possess the symmetry $\rho \rightarrow -\rho$ and $S_\alpha \rightarrow - S_\alpha$ we can truncate the system to the so-called ``reduced" system, in which $W_\nu$ and $B$ are independent of $\rho,$ while $S_\alpha$ is linear in $\rho.$ In this paper we consider the reduced system.

\subsec{Vacuum solutions}

Here we consider vacuum solutions of the equations of motion \eqIIe. The fields $B, W$ and $S_\alpha$ of the vacuum solution are denoted by $B^{(0)}, W^{(0)}$ and $S^{(0)}_\alpha,$ respectively. In particular we take $B^{(0)}$ to be constant, i.e.
\eqn\eqIIla{ B^{(0)} \equiv \nu}
Plugging this ansatz into \eqIIe\ we obtain the following three equations
\eqn\eqIIlb{\eqalign{ d W^{(0)} &= W^{(0)} \star \wedge W^{(0)} \cr
d S^{(0)}_\alpha &= W^{(0)} \star S^{(0)}_\alpha - S^{(0)}_\alpha \star W^{(0)} \cr
S^{(0)}_\alpha S^{(0) \, \alpha} &= - 2 i (1+ \nu K)}}
Note that the other two equations of \eqIIe\ are automatically satisfied by the ansatz \eqIIla.

First, let us discuss the third equation of \eqIIlb. We already mentioned that $S_\alpha$ and therefore also $S^{(0)}_\alpha$ is linear in $\rho.$ For the case $\nu =0$ we can choose $S^{(0)}_\alpha = \rho z_\alpha$, cf. \eqIIi. For general $\nu,$ $S^{(0)}_\alpha$ can be given by
\eqn\eqIIlc{ S^{(0)}_\alpha = \rho \tilde{z}_\alpha}
where we have introduced new auxiliary twistor variables $\tilde{z}_\alpha$ and $\tilde{y}_\alpha$ which are also known as ``deformed oscillators,''
\eqn\eqIIld{\eqalign{ \tilde{z}_\alpha &= z_\alpha + \nu \, w_\alpha k \cr
\tilde{y}_\alpha &= y_\alpha + \nu \, w_\alpha \star K \cr
w_\alpha &= (z_\alpha + y_\alpha) \int_0^1 dt\, t e^{i t (zy)}}}
The deformed oscillators $\tilde{y}_\alpha$ and $\tilde{z}_\alpha$ satisfy the commutation relations
\eqn\eqIIle{\eqalign{ [\tilde{y}_\alpha, \tilde{y}_\beta]_\star &= 2i \epsilon_{\alpha\beta} (1+\nu k) \cr
[\rho \tilde{z}_\alpha, \rho \tilde{z}_\beta]_\star &= -2i \epsilon_{\alpha\beta} (1+\nu K) \cr
[\rho \tilde{z}_\alpha, \tilde{y}_\beta]_\star &= 0}}
and therefore it is straightforward to verify that $S^{(0)}_\alpha,$ given by equation \eqIIlc, indeed satisfies the third equation of \eqIIlb.

As $d S^{(0)}_\alpha =0,$ the second equation of \eqIIlb\ implies that $W^{(0)}$ should commute with $S^{(0)}_\alpha,$ which according to the third line of \eqIIle\ can be achieved by taking $W^{(0)}$ to be independent of $k$ and $\tilde{z}_\alpha.$ Therefore $W^{(0)}$ is only a function of $x, \psi_{1,2}$ and $\tilde{y}$ and can be expanded as
\eqn\eqIIlf{ W^{(0)}(\tilde{y};\psi_{1,2}| x) = \sum\limits_{d,e=0}^1 \sum\limits_{n=0}^\infty {1\over{n!}} W^{\beta_1\dots\beta_n}_{00de}(x) \psi_1^d \psi_2^e \ \tilde{y}_{\beta_1} \star \dots \star \tilde{y}_{\beta_n}   }
It turns out that we only have to consider symmetric products in $\tilde{y}_{\beta_1} \star \dots \star \tilde{y}_{\beta_n}.$ Moreover, here we will consider only products with an even number of $\tilde{y}.$ Under the star product the auxiliary variables $\tilde{y}_\alpha$ generate the three-dimensional higher spin algebra \hsl. To be more precise, a symmetric product $\tilde{y}_{\beta_1}\star \dots \star\tilde{y}_{\beta_{2n}}$ corresponds to a generator of \hsl\ with spin $n+1.$ In particular the generators $T_{\alpha\beta}$ of the SL(2) subalgebra are given by
\eqn\eqIIlg{ T_{\alpha\beta} = - {{i}\over{4}} \{ \tilde{y}_\alpha, \tilde{y}_\beta \}_\star }
Multiplying symmetrized even-degree polynomials in $\tilde{y},$ using the commutation relations as given in the first line of equation \eqIIle\ and finally projecting on $k = \mp 1,$ the commutation relation are those of \hsl\ with $\lambda = {{1}\over{2}} (1 \pm \nu).$ More details, including the \hsl\ the structure constants, can be found in appendix B.

Finally, let us consider the last equation of motion which we have to solve:
\eqn\eqIIlh{d W^{(0)} = W^{(0)} \star \wedge W^{(0)}.}
This equation can be written as a flatness condition of a Chern-Simons theory with gauge group \hsl$\oplus$\hsl\ . In order to see this we introduce \hsl-valued gauge fields $A$ and $\Ab$ by
\eqn\eqIIli{ W^{(0)} = - {\cal P}_+ A - {\cal P}_- \Ab}
where $A$ and $\Ab$ are functions of $x$ and $\tilde{y}.$ We have introduced projection operators
\eqn\eqIIlj{ {\cal P}_\pm = {{1\pm\psi_1}\over{2}}}
obeying
\eqn\eqIIlk{ {\cal P}_\pm \psi_1 = \psi_1 {\cal P}_\pm = \pm {\cal P}_\pm, \qquad {\cal P}_\pm \psi_2 = \psi_2 {\cal P}_\mp}
Using \eqIIli\ we can rewrite \eqIIlh\ in the following form
\eqn\eqIIll{ dA + A\wedge\star A =0 \, , \qquad d\Ab + \Ab \wedge\star \Ab =0 }
which is precisely equation \intl. Therefore the equations of motion for $W^{(0)}$ are equivalent to flatness conditions of gauge fields $A$ and $\Ab$ defined by equation $\eqIIli.$

\subsec{Matter equations}
Let us now linearize the equations of motion \eqIIe\ around the vacuum solution constructed in the last section. In particular, in this paper we are interested in fluctuations of the field $B$ around the constant background $B^{(0)} = \nu.$ The fluctuations of $B$ are denoted by ${\cal C},$ i.e.
\eqn\eqIIma{B = \nu + {\cal C}}
For $W$ and $S_\alpha$ we do not consider any fluctuations. Substituting this ansatz \eqIIma\ into the equations of motion \eqIIe\ and using the equations \eqIIlb, we obtain two non-trivial equations for ${\cal C}$
\eqn\eqIImb{\eqalign{ d {\cal C} - W^{(0)} \star {\cal C} + {\cal C} \star W^{(0)} &=0 \cr
\left[ S^{(0)}_\alpha, {\cal C} \right]_\star &=0}}
Since $S^{(0)}_\alpha$ is given by equation \eqIIlc\ we can satisfy the second line of equation \eqIImb\ by demanding that ${\cal C}$ does not depend on $\tilde{z}_\alpha$ nor on $k.$ Therefore ${\cal C}$ is only a function of $\tilde{y}_\alpha, \psi_{1,2}$ and $x.$ Typically, ${\cal C}$ is decomposed as
\eqn\eqIImc{ {\cal C}(\tilde{y};\psi_{1,2} | x) = {\cal C}^{aux}(\tilde{y};\psi_{1} | x)  + {\cal C}^{dyn} (\tilde{y};\psi_{1} | x) \psi_2}
It turns out \ProkushkinBQ\ that ${\cal C}^{aux}$ gives rise to an auxiliary set of fields that can be set to zero consistently. By abuse of notation, we will use ${\cal C}$ instead of ${\cal C}^{dyn}$ to simplify the notation. Moreover we will decompose ${\cal C}$ under the projection operators ${\cal P}_\pm$ as given in equation \eqIIlj
\eqn\eqIImd{\eqalign{ {\cal C} = C (\tilde{y}| x) \, \psi_2 + \Ct (\tilde{y}| x) \, \psi_2 \cr }}
where $C = {\cal P}_+ {\cal C}$ and $\Ct = {\cal P}_- {\cal C}.$ If we also express $W^{(0)}$ in terms of gauge fields, eq. \eqIIli, we can rewrite the second line of \eqIImb\ in the form
\eqn\eqIIme{\eqalign{ dC + A \star C - C \star \Ab = 0 \cr
d\Ct + \Ab \star \Ct - \Ct \star A = 0}}
These are the equations of linearized matter interacting with an arbitrary higher spin background.

Comparing the first and second line of equation \eqIIme\ we see that the equations of motion for $C$ and $\Ct$ are related to each other by exchanging $A$ and $\Ab.$ Note that $A$ and $\Ab$ can be written in terms of the generalized vielbein $e$ and generalized spin connection $\omega,$
\eqn\eqIImea{ A = \omega + e \qquad \Ab =\omega - e}
Under the exchange of $A$ and $\Ab$ the generalized vielbein $e$ is odd. Therefore the sign of the generalized vielbein and hence the sign of any metric-like tensor field of odd spin is flipped under this operation.

\appendix{B}{\hsl, and Moyal vs. lone-star products  }

The \hsl\ structure constants are
\eqn\bc{g_u^{st}(m,n;\lambda) = {q^{u-2}\over2(u-1)!}\phi_u^{st}(\lambda)N_u^{st}(m,n)}
where
\eqn\bd{\eqalign{N_u^{st}(m,n) &= \sum_{k=0}^{u-1}(-1)^k
\left(\matrix{u-1 \cr k}\right)
[s-1+m]_{u-1-k}[s-1-m]_k[t-1+n]_k[t-1-n]_{u-1-k}\cr
\phi_u^{st}(\lambda) &= \ _4F_3\left[\matrix{\half + \lambda ~,~   \half - \lambda  ~,~ {2-u\over 2} ~ ,~ {1-u\over 2}\cr
{3\over 2}-s ~ , ~~ {3\over 2} -t~ ,~~ \half + s+t-u}\Bigg|1\right]\cr}}
We make use of the descending Pochhammer symbol,

\eqn\hswee{[a]_n  = a(a-1)...(a-n+1)}
$q$ is a normalization constant that can be scaled away by taking $V^s_m \rar q^{s-2}V^s_m$. As in much of the existing literature, we choose to set $q=1/4$.\vs

Recall the deformed oscillator commutation relations:
\eqn\abb{[\yta,\ytb]_\star=2i\epsilon_{\a\beta}(1+\nu k)}
which in our conventions $\epsilon_{12}=\epsilon^{12}=1$ is
\eqn\ac{[\yt_1,\yt_2]_\star=2i(1+\nu k)}
The beauty of the deformed oscillators is that under the action of the Moyal product\foot{In what follows, every product of $\yt$ is implicitly a Moyal product.}, their star commutator gives the oscillator algebra \abb.

To compute the Moyal product of two symmetric, even-degree oscillator polynomials, one uses \abb\ to symmetrize the product. To compute the lone-star product of two \hsl\ generators, one plugs into formula \gb. The purpose of this section is to provide evidence that these two multiplications are isomorphic upon identifying the map between the generator and polynomial bases, and using the relation
\eqn\aca{\l={1-\nu k\over 2}}
where $k^2=1$ is the Clifford element defined in section 2 and in appendix A. To our knowledge, this has not been proven in the literature.

Let us begin with the \slt\ subalgebra spanned by symmetric polynomials
\eqn\ad{S_{\a\b} = \yt_{(\a}\yt_{\beta)}}
These obey commutation relations
\eqn\af{\eqalign{[S_{11},S_{22}] = 8iS_{12}\cr
[S_{11},S_{12}] = 4iS_{11}\cr
[S_{12},S_{22}] = 4iS_{22}\cr}}
Comparing with the \slt\ subalgebra of \hsl\ canonically normalized as
 \eqn\ag{\eqalign{[V^2_1,V^2_{-1}] &= 2V^2_0\cr
[V^2_1,V^2_{0}] &= V^2_1\cr
[V^2_0,V^2_{-1}] &= V^2_{-1}\cr}}
these are equivalent under the assignment
\eqn\ah{\eqalign{V^2_1 &= \left({-i\over 4}\right)S_{11}~, ~~ V^2_0 = \left({-i\over 4}\right)S_{12}~, ~~V^2_{-1} = \left({-i\over 4}\right)S_{22} \cr}}

Having fixed \ah\ we can compare, on the one hand, the Moyal product of two $S_{\a\b}$, and on the other, the lone-star product between two of the $V^2_m$, and so on for higher spins. The work comes in symmetrizing the oscillator products, through tedious but straightforward application of \abb. We present results through spin-4:
\eqn\aii{\eqalign{\yto\ytt &= S_{12}+i(\nu k+1)\cr
\yto\yto\yto\ytt &= S_{1112} + i(\nu k +3)S_{11} \cr
\yto\yto\ytt\ytt &= S_{1122} + 4iS_{12} + {2\over 3}(\nu k +1)(\nu k -3) \cr
\yto\yto\yto\yto\yto\ytt &= S_{111112} + i(\nu k +5)S_{1111}\cr
\yto\yto\yto\yto\ytt\ytt &= S_{111122} + 8iS_{1112} + {4\over 5}(\nu k+3)(\nu k -5)S_{11}\cr
\yto\yto\yto\ytt\ytt\ytt &= S_{111222} + i(\nu k +9)S_{1122} + {2\over 5}(\nu k +3)(\nu k -15)S_{12}\cr &+ {2i\over 3}(\nu k+3)(\nu k+1)(\nu k-3)\cr
}}
All remaining products can be found by commutation or taking the adjoint, $\yto \leftrightarrow \ytt, i \rar -i$.

Using these, one can show by explicit computation that, at least through spin-4, any Moyal product of oscillator polynomials maps to the lone-star product of \hsl\ generators upon making the identification
\eqn\ao{V^s_m = \left({-i\over 4}\right)^{s-1}S^s_m}
where $S^s_m$ is the symmetrized product of $2s-2$ oscillators,
with $m$ defined as
\eqn\ai{2m=N_1-N_2}
The prefactor depends on the \hsl\ normalization factor $q=1/4$ which we have been using.

As a first check, \ah\ along with the fact that
\eqn\aij{(V^2_{\pm1})^{s-1} = V^s_{\pm(s-1)}}
implies that this is trivially true for elements with $m=\pm(s-1)$.

Let us demonstrate this equivalence for a few examples. Using \aii\ one obtains the Moyal products of spin-2 polynomials:
\eqn\ak{\eqalign{
S_{11}\star S_{11} &= S_{1111}\cr
S_{22}\star S_{22} &= S_{2222}\cr
S_{12}\star S_{11} &= S_{1112}-2iS_{11}\cr
S_{12}\star S_{22} &= S_{1222}+2iS_{22}\cr
S_{11} \star S_{22} &= S_{1122}+4iS_{12}+{2\over 3}(\nu k+1)(\nu k-3)\cr
S_{12}\star S_{12} &= S_{1122}-{1\over 3}(\nu k+1)(\nu k-3)\cr}}

We now compare this to the lone-star products of spin-2 generators:
\eqn\al{\eqalign{
V^2_1\star V^2_1 &= V^3_2\cr
V^2_{-1}\star V^2_{-1} &= V^3_{-2}\cr
V^2_0\star V^2_1 &= V^3_1-{1\over 2}V^2_1\cr
V^2_0\star V^2_{-1} &= V^3_{-1}+{1\over 2}V^2_{-1}\cr
V^2_1\star V^2_{-1} &= V^3_0+V^2_0-\left({\l^2-1\over 6}\right)\cr
V^2_0\star V^2_0 &= V^3_0+{\l^2-1\over 12}\cr}}
These are isomorphic under \aca\ and \ao.

A less trivial example is the product
\eqn\amm{\eqalign{S_{1112}\star S_{22} &= {1\over 4}(\yto\yto\yto\ytt + \yto\yto\ytt\yto + \yto\ytt\yto\yto + \ytt\yto\yto\yto) \ytt\ytt\cr
&= S_{111222} + 6iS_{1112} + {2\over 5}(\nu k +3)(\nu k -5)S_{12}\cr}}
Compare to the lone-star product
\eqn\amp{V^3_1\star V^2_{-1} = V^4_0 + {3\over 2}V^3_0 -\left({\l^2-4\over 10}\right)V^2_0}
These are isomorphic under \aca\ and \ao.

We conjecture that the isomorphism is valid for all spins under the identification \ao.

\appendix{C}{Derivation of (5.14)}

Given
\eqn\pkbs{ L_0 = -{1\over 4} (x\p_x + \p_x x)   ~,\quad L_{-1} = {1\over 2}  \p_x^2~,\quad L_1  = {1\over 2}  x^2 }
and

\eqn\pkbt{ V^s_m = (-1)^{m-1} {(m+s-1)!\over (2s-2)!}\underbrace{ \big[ L_{-1}, \ldots [L_{-1}, [L_{-1},}_{s-1-m} L_1^{s-1}]] \big]}
we need to show
\eqn\pkbu{ V^s_0 x^{-2} =  {(-1)^s[(s-1)!]^2 (2s-1)!!  \over 2^{s-1}(2s-2)!} x^{-2}  }

We start from
\eqn\pkbv{ e^{tL_{-1}} f(x) e^{-tL_{-1}} = f(x+t\p_x)}
which follows by thinking of $L_{-1}$ as the Hamiltonian for a free particle.  In particular, this implies
\eqn\pkbw{  e^{tL_{-1}} (L_1)^{s-1}  e^{-tL_{-1}} = {1\over 2^{s-1}} (x+t\p_x)^{2s-2} }
Expanding the left hand side in powers of $t$, the desired term yielding $V^s_0$ is the $t^{s-1}$ term.   This term preserves the power of $x$, and so we can write
\eqn\pkbx{V^s_0 x^{-2}  = -\left[ { [(s-1)!]^2  \over 2^{s-1}(2s-2)!} (x+\p_x)^{2s-2}x^{-2}\right] \Bigg|_{ x^{-2}~ {\rm term} }  }
To extract the $x^{-2}$ term on the right hand side we write a contour integral and integrate by parts,
\eqn\pkby{  (x+\p_x)^{2s-2}x^{-2} \Big|_{{\rm coeff~of~} x^{-2}}= {1\over 2\pi i} \oint\! dz z (z+\p)^{2s-2} {1\over z^2} ={1\over 2\pi i} \oint\! dz {1\over z^2}  (z-\p)^{2s-2} z }
Now write  $(z-\p)^{2s-2}$ as the $t^{2s-2}$ term in  $e^{t(z-\p)}=e^{{-t^2/2}} e^{tz}e^{-t\p}$ and perform the integral.  This yields
\eqn\pkbz{ (x+\p_x)^{2s-2}x^{-2} \Big|_{{\rm coeff.~of}~ x^{-2}} = (-1)^{s-1}(2s-1)!!}
Plugging this result into \pkbx\ yields our desired formula \pkbu.

\appendix{D}{Low spin results: $C^s_0$ in AdS}
We present the explicit formulae for the components $C^s_0$ of the master field in AdS, through $s=8$, obtained by recursive solution of the $V^s_{0,\rho}$ equations in AdS:
\eqn\apda{\pr C^s_0 + 2C^{s-1}_0+ C^{s+1}_0g^{(s+1)2}_3(0,0)=0}
These can be compared to the spin $s$ formula \gt. We use the temporary notation
\eqn\gea{g^{s2}_3(0,0)\equiv g^s}
and the following expressions act on $C^1_0$:
\eqn\gf{\eqalign{C^2_0 &= -(g^2)^{-1}\pr\cr
C^3_0 &= (g^2g^3)^{-1}\cdot(\pr^2-2g^2)\cr
C^4_0 &= (g^2g^3g^4)^{-1}\cdot\left(-\pr^3 + 2(g^2+g^3)\pr\right)\cr
C^5_0 &= (g^2g^3g^4g^5)^{-1}\cdot\left(\pr^4-2(g^2+g^3+g^4)\pr^2+4g^2g^4\right)\cr
C^6_0 &= (g^2g^3g^4g^5g^6)^{-1}\cdot\left(-\pr^5 + 2(g^2+g^3+g^4+g^5)\pr^3-4(g^5(g^3+g^2)+g^4g^2)\pr\right)\cr
C^7_0 &=(g^2g^3g^4g^5g^6g^7)^{-1}\cdot\Big( \pr^6-2(g^2+g^3+g^4+g^5+g^6)\pr^4\cr&+4(g^6(g^4+g^3+g^2)+g^5(g^3+g^2)+g^4g^2)\pr^2-8g^2g^4g^6\Big)\cr
C^8_0 &=(g^2g^3g^4g^5g^6g^7g^8)^{-1}\cdot\Big( -\pr^7+2(g^2+g^3+g^4+g^5+g^6+g^7)\pr^5\cr&-4(g^7(g^5+g^4+g^3+g^2)+g^6(g^4+g^3+g^2)+g^5(g^3+g^2)+g^4g^2)\pr^3\cr&+8(g^7(g^5(g^3+g^2)+g^4g^2)+g^6g^4g^2 )\Big)\pr\cr}}

\listrefs\end